\documentclass[prd,twocolumn,showpacs,preprintnumbers,amsmath,amssymb]{revtex4}

\def\kms{\mbox{km\,s$^{-1}$}}

\def\s{\,{\rm s}}
\def\K{\,{\rm K}}

\def\lya{\rm{Ly\alpha}}

\def\rtwo{r_{_{\rm 200}}}
\def\ub{{\bf u}_{_{\rm bar}}}
\def\dub{{\dot {\bf u}}_{_{\rm bar}}}
\def\ud{{\bf u}_{_{\rm dm}}}

\def\rfive{r_{_{\rm 500}}}

\def\cnfw{{C_{_{\rm NFW}}}}

\def\mdm{{M_{_{\rm h}}}}
\def\mfin{{M_{_{\rm final}}}}

\def\rhoc{{\rho_{_{\rm c}}}}

\def\rs{{r_{_{\rm s}}}}

\def\lcdm{{\rm{ \Lambda CDM}}}

\def\hms{{ \rm{ \,h^{-1}M_\odot} }}
\def\kms{{\rm{ \, km \, s^{-1}}}}
\def\la{\lower.5ex\hbox{$ \; \buildrel < \over \sim \; $}}
\def\ga{\lower.5ex\hbox{$ \; \buildrel > \over \sim \; $}}
\usepackage{epsfig}

\def\hmpc{\, { {\rm h}^{-1} {\rm Mpc} }}
\def\rhmpc{\, { {\rm {\rm h}\ Mpc}^{-1} }}

\def\mpc{\, { {\rm Mpc} }}

\newcommand{\phiv}{\phi_v}
\newcommand{\vg}{\mathbf{v}_{_{\rm bar}}}
\newcommand{\dvg}{\dot{\mathbf{v}} _{_{\rm bar}}}
\newcommand{\pot}{\phi_{_{\rm grav}}}
\newcommand{\phidm}{\phi_{_{\rm dm}}}
\newcommand{\phibar}{\phi_{_{\rm bar}}}
\newcommand{\phis}{\phi_{_{\rm s}}}
\newcommand{\nab}{\mathbf{\nabla}}
\newcommand{\dg}{\delta_{_{{\rm bar}}}}

\newcommand{\dm}{\delta_{_{\rm dm}}}

\newcommand{\ddg}{\dot\delta _{_{{\rm bar}}}}

\newcommand{\ddm}{\dot\delta _{_{{\rm dm}}}}
\newcommand{\dddg}{\ddot\delta _{_{{\rm bar}}}}

\newcommand{\dddm}{\ddot\delta _{_{{\rm dm}}}}
\newcommand{\bk}{\mathbf{k}}
\newcommand{\rdm}{\rho _{_{\bf dm}}}
\newcommand{\rbar}{\rho _{_{\bf bar}}}
\newcommand{\mrdm}{\bar{\rho} _{_{{\rm dm}}}}
\newcommand{\om}{\Omega _{_{\rm m}}}
\newcommand{\omz}{\Omega _{_{{\rm m0}}}}
\newcommand{\olm}{\Omega _{_{\Lambda}}}
\newcommand{\olz}{\Omega _{_{\Lambda0}}}
\newcommand{\omg}{\Omega _{_{{\rm bar}}}}
\newcommand{\omdm}{\Omega _{_{{\rm dm}}}}

\newcommand{\msat}{M_{_{\rm 1}}}

\newcommand{\ace}{{ReBEL }}
\newcommand{\acec}{{ReBEL}}

\newcommand{\beqa}{\begin{eqnarray}}
\newcommand{\eeqa}{\end{eqnarray}}

\begin{document}
%ReBEL: daRk Breaking Equivalence principLe

\title{Cosmology with Equivalence Principle Breaking in the Dark Sector}

\author{Jose Ariel Keselman}
\email{kari@tx.technion.ac.il}
\affiliation{Physics department, Technion, Haifa 32000, Israel}
\author{Adi Nusser} \affiliation{Physics Department and the Asher
    Space Research Institute, Technion, Haifa 32000, Israel}
\author{P.~J.~E. Peebles} \affiliation{Joseph Henry Laboratories,
    Princeton University, Princeton, NJ 08544, USA}

\begin{abstract}
A long-range force acting only between nonbaryonic particles would be
associated with a large violation of the weak equivalence principle. We
explore cosmological consequences of this idea, which we label
ReBEL (daRk Breaking Equivalence principLe).  A high resolution
hydrodynamical simulation of the distributions of baryons and dark matter confirms our
previous findings that a \ace force of comparable strength to
gravity on comoving scales of about $ 1\hmpc $ causes voids between
the concentrations of large galaxies to be more nearly empty,
suppresses accretion of intergalactic matter onto galaxies at low
redshift, and produces an early generation of dense dark matter
halos.  A preliminary analysis indicates the \ace scenario is consistent with the one-dimensional
power spectrum of the Lyman-Alpha forest and the three-dimensional galaxy
auto-correlation function. Segregation of baryons and DM in galaxies and systems of galaxies is
a strong prediction of ReBEL. \ace naturally correlates the baryon mass
fraction in groups and clusters of galaxies  with the system mass, in agreement
with recent measurements.
\end{abstract}

\pacs{98.80.-k, %Cosmology
11.25.-w, %Strings and branes
95.35.+d,
% Dark matter (stellar, interstellar, galactic, and cosmological)
98.65.Dx
%Superclusters; large-scale structure of the Universe (including voids, pancakes, great wall, etc.)
%More PACS available at http://publish.aps.org/PACS/pacsgen.html
}

\maketitle

\section{INTRODUCTION}%
\label{sec:introduction}

The $ \lcdm $ cosmology (the relativistic hot Friedmann-Lema\^\i tre model with a cosmological constant
and Cold Dark Matter, DM) is very successful at matching
the large-scale distributions of matter and radiation, including
 the angular Power Spectrum (PS) and polarization
of the cosmic microwave background  and the galaxy correlation
functions and PS, if one allows for mild biasing between the mass
and galaxy distributions. But there are problems with structure formation, and  there
is a possible remedy, \acec, a long-range force of attraction operating only on the DM.
This has been studied in numerical simulations of structure formation in
\cite{nusser05, kk06, hellwing08, keselman09}. Here we present new results on structure
formation on cosmological length scales from joint modeling of baryons and DM using the
Smoothed-Particle Hydrodynamics (SPH) method and the DM TreePM code Gadget2 \cite{springel05},
modified to take account of the added force on the DM. Our two simulations compare evolution
of structure with and without \acec. We also use the Halo Occupation Distribution (HOD)
framework to compare model predictions to  statistical measures of large-scale structure derived from
recent galaxy surveys.

We consider \ace in the form of an attractive force between DM particles alone
of the form
\begin{equation}
    \label{eq:eq1} {\bf F}=-\frac{\beta Gm^2}{a^2r^2}e^{-r/\rs}\left( \frac{1}
    {r}+\frac{1}{\rs} \right)\bf{{r}},
\end{equation}
as in \cite{farrar04, nusser05, keselman09}. Here $a \bf{r} $ is the
physical separation vector, where $a(t)$ is the cosmological
expansion parameter. The screening length, here set to $\rs=1\hmpc$ (where the Hubble
parameter is $ H_0=100h\kms\mpc^{-1}$), is
constant in comoving coordinates. The Newtonian gravitational
constant is $ G $ and the DM particle mass is $ m $.

Frieman and Gradwohl pointed out the constant $\beta$ that measures the strength of the
\ace force is bounded above by its tendency
to separate stars from DM \cite{frieman93},
contrary to the observation that some
dwarf spheroidal companions of the Milky Way have
retained their DM halos.
Kesden \& Kamionkowsky \cite{kk06} made the
excellent point that a moderately
strong \acec, with $\beta\sim 0.1$, separates stars from DM in a satellite galaxy in one-sided stream. This is quite contrary to the observed
classical double stream of stars in the Sagittarius
galaxy \cite{ibata94}. We consider here the stronger \ace force
$\beta=1$. This passes the Frieman-Gradwohl test in the cases considered in \cite{keselman09}. It also passes the Kesden-Kamionkowsky
test because \ace would pull the Sagittarius stars away
from the DM halo before close passages by the Milky Way
tidally disrupt the DM-free but still gravitationally bound star
cluster in the observed two streams. The process is illustrated
in Figures $3$ and $11$ in Keselman et. al. \cite{keselman09}.
Bean et al. \cite{bean08} show that $\beta\simeq 1$ also so far
passes the constraints from the cosmological tests.

Section \ref{sec:sec_issues} reviews issues that motivate analysis
of observational consequences of \acec, with a guide to the issues
studied in the present paper. The numerical
modelling is described in \S~\ref{sec:sec_numerics}.  Results are
presented in \S~\ref{sec:sec_results}, beginning with
a general description of the main
results. The more elaborate analyses are divided into several
subsections describing the different algorithms used for finding halos
and voids and the method used to calculate the $ \lya $ PS.
A concluding discussion is presented in \S~\ref{sec:sec_discussion}.

\section{Issues and motivations}%
\label{sec:sec_issues}

We review here our selection of apparent challenges to structure formation
 that motivate consideration of the addition of \ace to $\lcdm$.
 For another overview of the situation see \cite{peri08}.

The $ \lcdm $ cosmology tends to over-populate halos of galaxies with satellites and voids with dwarf galaxies.
The former \cite{klypin99, moore01,
madau08, simon07, strigari07} may be resolved by the effect of plasma  pressure or winds that leave low mass DM halos with too few baryons to be observable. \ace offers another stripping effect, the differential acceleration of baryons and DM \cite{keselman09}, and \ace may also promote more complete merging of small halos \cite{nusser05, hellwing08}.
The latter issue \cite{peebles01, gott03, goldberg04} is best expressed as a comparison of counts of galaxies actually observed
in denser regions to what is seen in voids
\cite{peebles07, tik09}. In particular,
the nearby Local Void contains just two known galaxies
\cite{karach04, meyer04, PeeblesNusser10}, while scaling from counts of galaxies observed elsewhere would predict the presence of about 15 to 30
galaxies with $ -18\la M_{_{\rm B}}\la -10 $ in the Local Void.
The possible role of a \ace force in more completely emptying
voids is explored in \cite{keselman09} and in \S~\ref{sub:sub_voids}.

While $ \lcdm$ correctly predicts the
general density profile of clusters
of galaxies it tends to underestimate the concentration
parameter, resulting in profiles that are too shallow
with respect to observations based
on strong and weak lensing
\cite{broadhurst05, umetsu08, broadhurst08, %broadhurst05b,
broadhurst05} and 
X-Ray emission \cite{lemze08}.
We show in \S~\ref{sec:halomassprofile} that \ace increases the concentration by earlier assembly of clusters, which is in the direction indicated by the observations.

The $\lcdm$ cosmology underpredicts
the abundance of disk-dominated galaxies \cite{stewart08, mayer08}.
This is because merging accretion of extragalactic debris by
DM halos of mass $ 10^{11}-10^{13}M_\odot $ is predicted to
continue to modest redshifts, causing thin disks that formed
earlier to thicken into bulges by gravitational heating \cite{weinmann06, wyse01, kormendy05,
wright09}. The predicted merging also is difficult to reconcile
with the insensitivity to environment
of the correlations among luminosity, radius, velocity
dispersion and color in early-type and late-type
galaxies  \cite{hogg04, bernardi06, disney08, vandenB10}.
 We show in \S~\ref{sec:halomerginghistory} that \ace promotes
 significantly earlier assembly of galaxies, which is in the
 direction wanted to relieve these problems.

The earlier structure formation under \ace would promote earlier reionization. 
Polarization measurements of the cosmic microwave background
radiation indicate reionization of hydrogen by redshift
$11$, implying early structure formation \cite{cen06, kogut03}. Galaxy/star
formation in the $\lcdm $ with $\sigma_8\approx 0.9$ could be
early enough to account for such  early reionization. However,
extreme efficiency of UV photon production by early structures is
needed for $\sigma_8\approx 0.8 $, the value favored by  WMAP data
\cite{benson06}. \ace may relieve this condition.

Earlier merging would lead to stronger clustering of more
massive galaxies now and the clustering of the intergalactic medium at
redshift $z\sim 3$ observed as the Lyman-Alpha forest.
The galaxy two-point correlation function and PS found
in \cite{hellwing08} in and here in \S~\ref{sec:clusteringofgalaxies}
seem to be consistent with what is observed within the usual uncertainties
of biasing. Dark matter halos are more massive
in \ace (\S~\ref{sec:halomassfunction}), and the
difference from the standard model is more pronounced
at redshift $z=1$ than the present. Remaining to be
explored is the effect on the galaxy luminosity function.
We conclude in \S~\ref{sec:lyaforest} that the model
parameters for the intergalactic medium at $z\sim 3$ 
can be adjusted within the available constraints
to fit the one-dimensional $\lya$ PS about equally well
in the standard model and in \ace with $\beta=1$.

Groups of galaxies in which the virial temperature is large enough that
intragroup plasma would  be expected to be detectable in X-ray emission
are observed to contain a significantly smaller baryon mass fractions than
the cosmic mean (e.g. \cite{afshordi07, hoekstra05, mcg08, mcg09, mccarthy07}).
This may be a result of the tendency of \ace to segregate DM from baryonic
matter, as shown in \cite{baldi08, baldi09} and further analyzed
in \S~\ref{sec:baryonmassfraction}.

The more rapid structure formation that seems to be indicated
by the phenomenology we have reviewed could follow from at
least three adjustments  of $\lcdm$: a departure from near
scale-invariant adiabatic Gaussian initial conditions,
a modified gravity with a departure of the inverse square law 
that preserves the weak equivalence principle \cite{SilvestriTrodden}, 
or, in general relativity theory, a long-range force that 
operates only on the DM particles. We consider this last option.
Superstrings offer scenarios for such a force in the dark sector
\cite{krip88, bran89, gubser04}. The force would cause the
accelerations of cosmic objects to depend on their relative DM
content, in what may be termed a breaking  of the weak
equivalence principle, or \ace (daRk Breaking Equivalence
principLe). Bovy and Farrar \cite{bovy09} point out that if the DM
interacted with visible matter strongly enough to allow laboratory
DM detection then \ace could induced a fifth force in the visible
sector that is much stronger than allowed by the E\"otv\"os
laboratory constraint. It should be noted, however, that a
laboratory DM detection need not vitiate \acec, for there may be
several types of DM in addition to massive neutrinos, including
one with a significant interaction with the visible sector but not
\ace, and another that is coupled to \ace but not the visible
sector \cite{misner73, bovy09, carroll09}.

\begin{figure*}[htpb]
    \centerline{\epsfig{figure=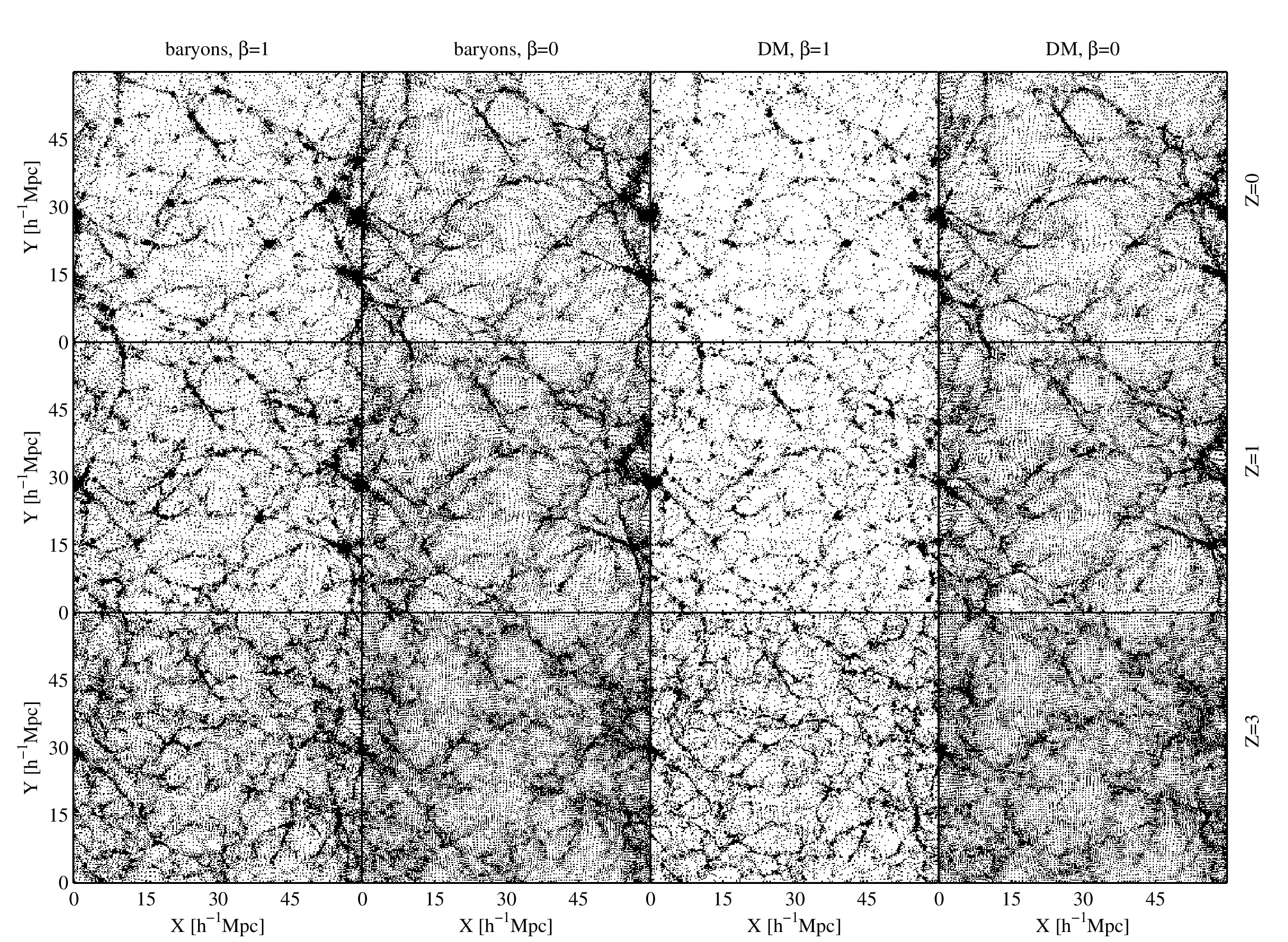, width=18.33cm,
    height=12.7cm}}
    \caption{Particle distributions in a slice $ 2\hmpc $ thick.
    Only a random subsample, containing $10\%$ of the particles is shown.}%
    \label{fig:particles}
\end{figure*}

\section{Numerical modelling}%
\label{sec:sec_numerics}

\subsection{The simulations} 
We have modified the Gadget2 code to include, in addition to gravity,
the force in Eq.~\ref{eq:eq1} operating between the DM
particles only. The simulations include gas (baryonic) particles
subject to pressure and gravitational forces but not the \ace
force.  The hydrodynamic equations are solved using the particle-based SPH method in Gadget2.
The modification is such that the error in the
calculation of the total force is the same as in the standard,
public Gadget2.  Appendix \ref{app:app_modification} describes the
details of the modification. The modified code is tested as described
in appendix \ref{app:app_tests}.

We obtain for comparison two simulations, with and without \acec.
The cosmological background is the
same in the two simulations, with density parameters $
\Omega_{DM}=0.23 $, $ \Omega_{BAR}=0.046 $, $ \Omega_\Lambda=0.72
$, with Hubble parameter $h=0.7$, all consistent with \cite{wmap5}. Encouraged by our previous
results \cite{nusser05, keselman09} we adopt the screening length $
r_s=1\hmpc $ and \ace force parameter $ \beta=1 $ in Eq.
\ref{eq:eq1}.

The simulation box is $ 60\hmpc $
on the side. It contains $ 128^3 $ DM particles and an equal
number of baryonic particles. The DM particles have mass $ 6.57
\times 10^9  \hms$, the baryon particles $ 1.31 \times 10^9 \hms $. The
bayonic equation of state is that of a perfect gas with $
\gamma=5/3 $, neglecting energy sinks and sources. The specific 
entropy is conserved except at shocks.  The initial entropy corresponds to a
temperature of 5\,K at the starting redshift of the simulation,
$z=1000$. Shock waves generated in collapsing structures heats the gas to
the much higher temperatures noted in the next section. 

Our two simulations start from identical initial conditions at
$ z=1000 $, when the \ace force is assumed to start operating \cite{gubser04}.
The initial conditions are generated from the linear PS of
the standard $\lcdm$ \cite{bbks86} cosmology for our choice of
parameters. We use the Zel'dovich
displacements of points and their peculiar velocities in an initially
uniform body-centered cubic lattice. The initial density fluctuations
are normalized so the linearly extrapolated (without \ace)
rms value of the density fluctuations at $ z=0 $ in spheres of radius
$ 8\hmpc $ is $ \sigma_8=0.8 $.

To identify DM halos in the simulation we apply the
``Friends-of-Friends'' (FOF) algorithm \cite{davis85}. The
interparticle linking length is $ b=0.2 $ in units of the mean
particle separation.

\subsection{Galaxies and Halos in the simulations}%
\label{sub:g}

To compare results from the simulations to
the observed large-scale distribution of galaxies we need
a recipe for identifying mock galaxies.  The
most thorough way is to trace halos in the
simulation forward in time and assign them galaxies according to
semi-analytic galaxy formation recipes (e.g. \cite{kauff97}),
but this requires simulations with larger resolution than
presented here. We use instead the simpler Halo Occupation
Distribution (HOD) framework \cite{krav04}, in which a DM halo with mass $ \mdm $ above a threshold $ M_{_{\rm min}} $ contains a central galaxy, and, above a larger threshold $M_1$, satellite galaxies.

We adopt HOD parameters such that galaxies in the simulations
represent the observed galaxies more luminous than $ {L_\star}/2.5 $,
corresponding to absolute magnitude $M_r=-19.5 $, with the observed
number density $ 0.01\, {\rm h}^3\, \rm{Mpc}^{-3} $ adopted from table 2 in \cite{zehavi05}.
In the conventional case, $\beta = 0$, the HOD parameters also produce a
satisfactory fit to the observed number density of voids above a mass
threshold of $ 10^{10} \hms $ and the distributions of galaxies  within
them \cite{tik09}. The $\beta=1$ case requires a modification of the HOD
prescription to fit the voids, as we will describe.

In the $\beta = 0$ case the number of satellite galaxies
assigned to a halo of mass $ \mdm $ is drawn from a Poison distribution
with mean
\begin{equation}
    \left\langle N \right\rangle = \left( \frac{\mdm}{\msat}\right)^\alpha
    \label{eq:mean_sat}\; .
\end{equation}
The three parameters $ M_{_{\rm min}} $, $ M_1 $, and $ \alpha $ are
fixed to match the galaxy number density and correlation function.
Each satellite is placed on a randomly chosen halo DM particle.

To match both the the properties of voids and the distribution of  bright galaxies in the $\beta=1$ case we must modify the HOD prescription. We adopt a
broken power-law model for the number of satellites in a halo,
\begin{equation}
    \left\langle N \right\rangle = \left\{%
    \begin{array}{ll}
        \left( \frac{\mdm}{\msat}\right)^\gamma &\mbox{ if $ \mdm<M_{_{\rm
        s}} $}, \\
        \left( \frac{M_{_{\rm s}}}{\msat}\right)^\gamma + \left( \frac{\mdm-M_
        {_{\rm s}}}{\msat}\right)^\alpha &\mbox{otherwise,}
    \end{array}
    \right.
\end{equation}
which is continuous at the mass scale $ M_{_{\rm s}} $.  Also, we need a more extended distribution of satellite galaxies than results from the $ \beta=0 $ prescription. We achieve this by increasing the
satellite distances from the halo center by a constant
multiplicative factor, $\zeta$,
relative to the DM distribution in the host halo. This produces a
satellite distribution
similar to that of the $\beta=0$ case \cite{man06,adami98}.

\section{Results}%
\label{sec:sec_results}

Fig. \ref{fig:particles} shows a
representative distribution of the gas and DM particles in a slice
$2\hmpc$ thick through the simulations at three
redshifts, as indicated in the figure.  Comparison between the DM
distributions (two columns to the right) with \ace ($ \beta=1 $)
and the standard model ($ \beta=0 $) clearly shows the \ace enhancement
of small-scale clustering of the DM:  voids are emptier
and the presence of individual clumps more pronounced.
There also is a stronger redshift evolution in the DM
particle distribution for $ \beta=1 $. (Here and throughout we use coordinates comoving with the general expansion.)  In the $ \beta=0 $
simulation the baryonic component closely follows the DM on
scales larger than the Jeans length, as expected. Also as expected is
the behavior for $\beta=1 $, where  the baryons are affected by the \ace force only indirectly, through the gravitational attraction of the more strongly clustered DM.  The differences
between the baryon distributions in the $ \beta=1 $ and $
\beta=0 $ simulations in the two columns to the left, though
important, are indeed less pronounced than in the DM
distributions.

\begin{figure}[htpb]
    \centerline{\epsfig{figure=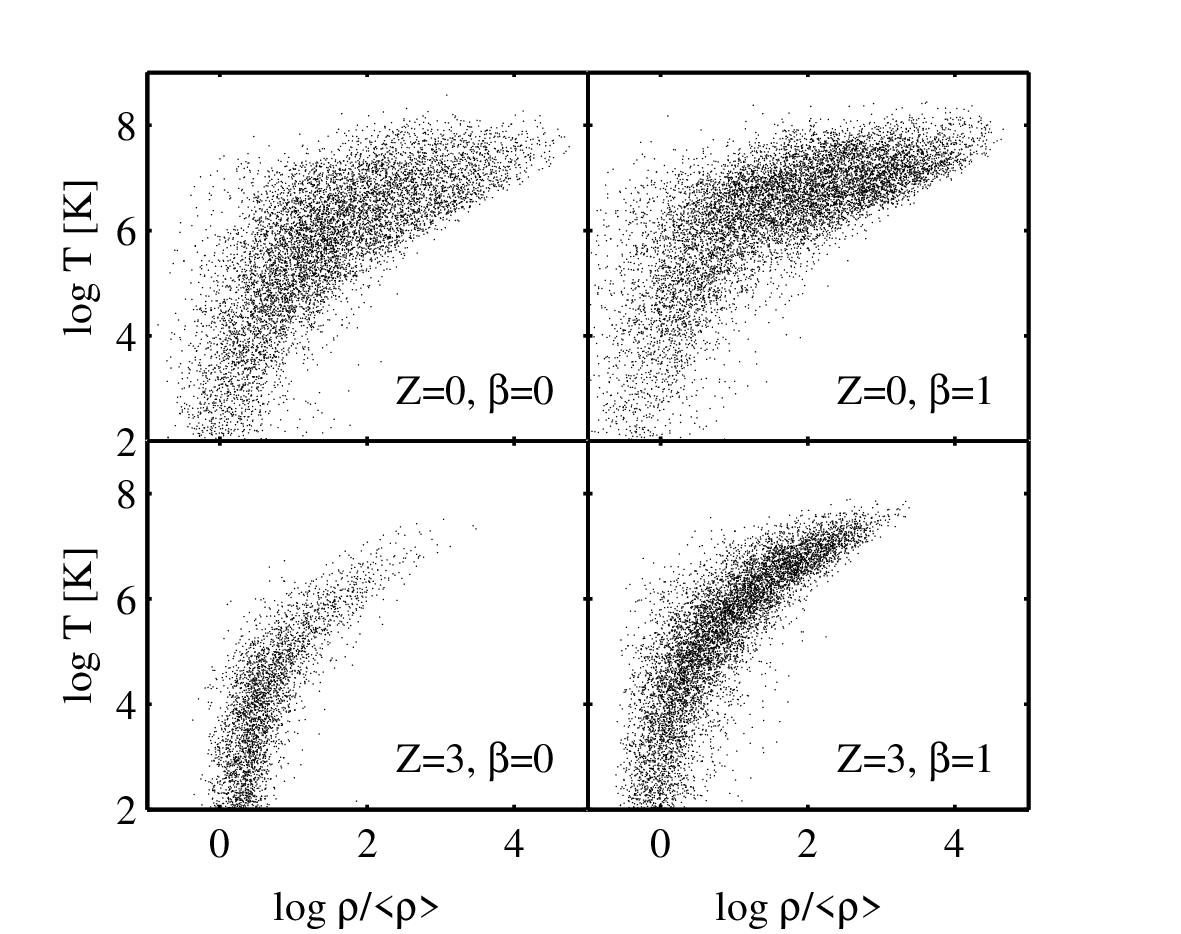,   width=8cm,
    height=6.3cm}}
    \caption{Scatter plot of temperature versus density of
    particles in the $\beta=0$ (left panels) and $\beta=1$ (right) simulations, 
     at redshifts $z=0$ (top panels) and $z=3$ (bottom)}
    \label{fig:temp_scat}
\end{figure}

Fig. \ref{fig:temp_scat} shows a scatter plot of temperatures and
overdensites of $10^4$ SPH gas particles  in the two simulations at
two redshifts.  The top-right corner of this $T-\rho$ plane is more
densely populated for $\beta=1$,
and the difference is more pronounced at $z=3$: in \ace more baryons
are in dense hot regions. Apart from that the diagrams for $\lcdm$ and
\ace are similar, consistent with the rough similarity of the baryon
distributions in Fig.~\ref{fig:particles}.

\subsection{Correlation Functions and Power Spectra}

\subsubsection{The distributions of baryons and dark matter}

\begin{figure}
    \centerline{\epsfig{figure=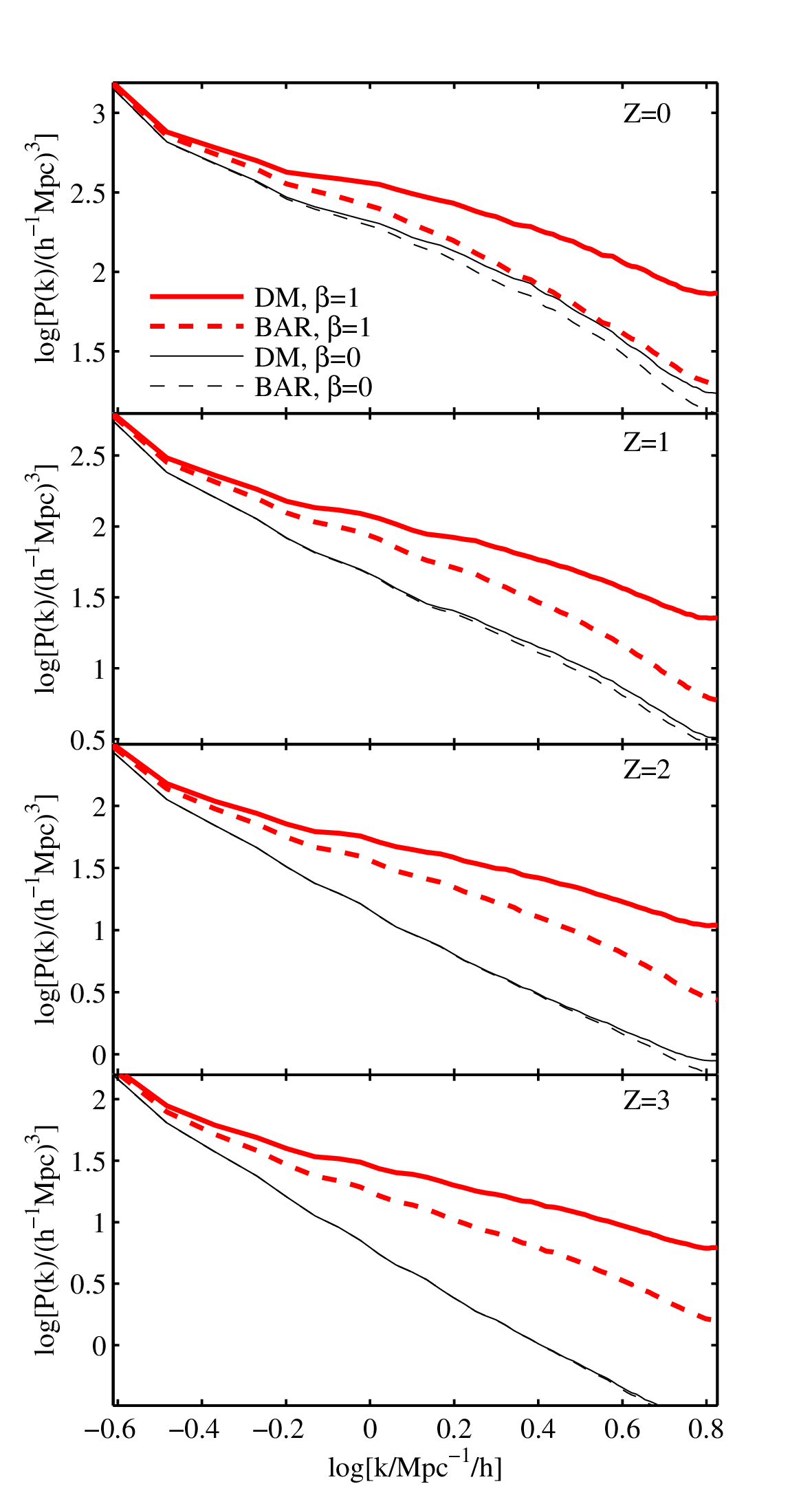, width=8cm,
    height=15.7cm}}
    \caption{PS of the baryon and DM distributions for $
    \beta=0 $ and $ \beta=1 $ at the redshifts indicated in
    the figure.}%
    \label{fig:power}
\end{figure}

\begin{figure}[htpb]
    \centerline{\epsfig{figure=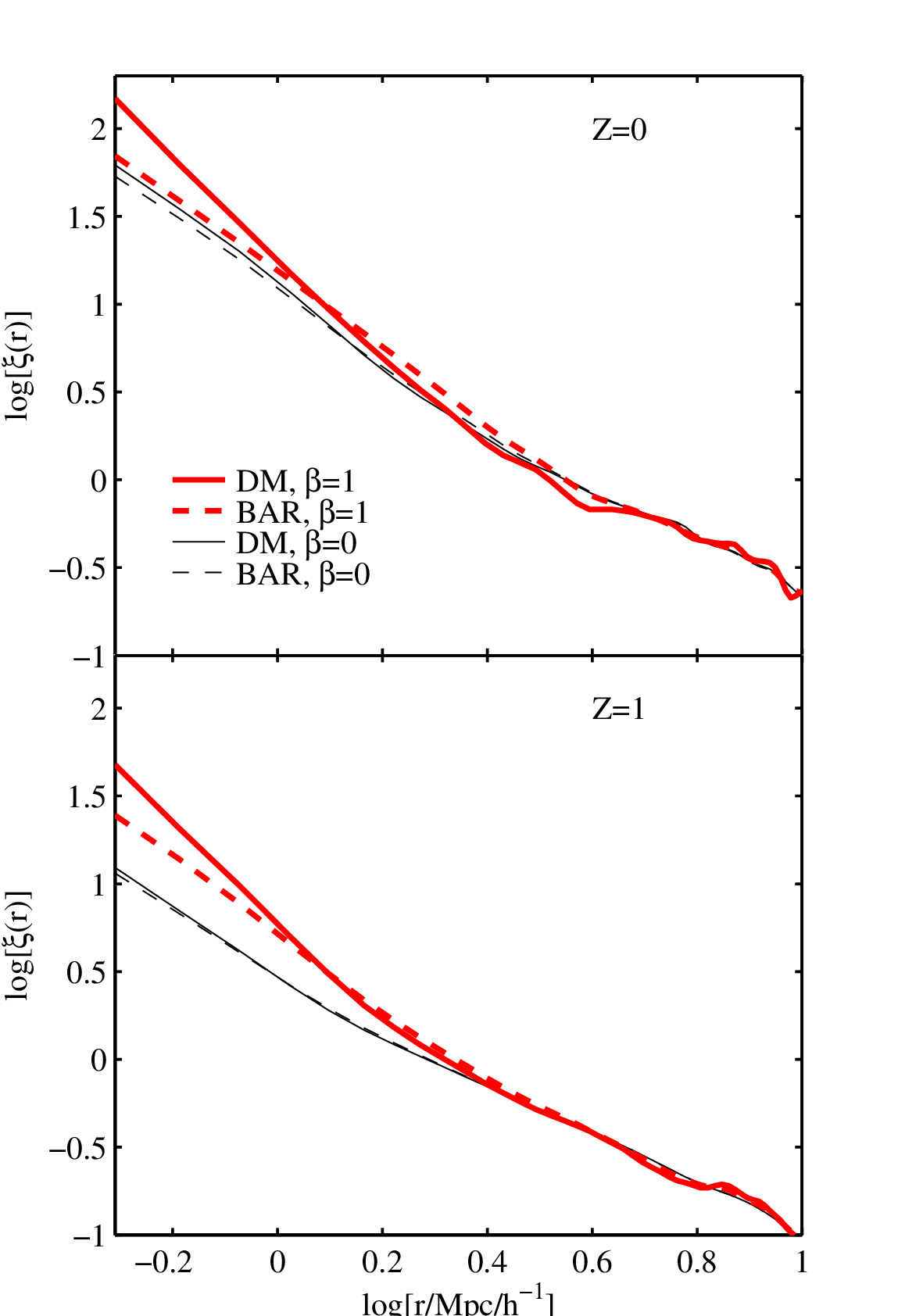, width=8cm,
    height=11.41cm}}
    \caption{Autocorrelation functions for the two components of
    the two simulations at the current epoch and redshift $z=1 $.}%
    \label{fig:corr}
\end{figure}

The clustering properties of the particle distributions measured by
their PS are shown in in Fig. \ref{fig:power} at several redshifts. Thin lines correspond to the standard scenario, $ \beta=0 $, while the thick lines show the effect of \ace with $ \beta=1 $.  As expected, the PS in the two simulations are almost the same on scales large compared to the screening length, $\rs=1\hmpc$, while \ace considerably increases the clustering of the DM on smaller scales. In the standard scenario the baryon
PS (thin dashed) closely follows that of the DM (thin
solid) at all redshifts, while in the \ace scenario the baryon PS
(thick dashed) falls below the DM curve (thick solid).  The baryons in the
\ace simulation are significantly more clustered than in the standard model
at higher redshifts, but at  $ z=0 $ the baryons in the standard and \ace
scenarios have relaxed to quite similar PS. This evolution is illustrated
in another way by the autocorrelation functions in Fig. \ref{fig:corr}.

\begin{figure}[htpb]
    \centerline{\epsfig{figure=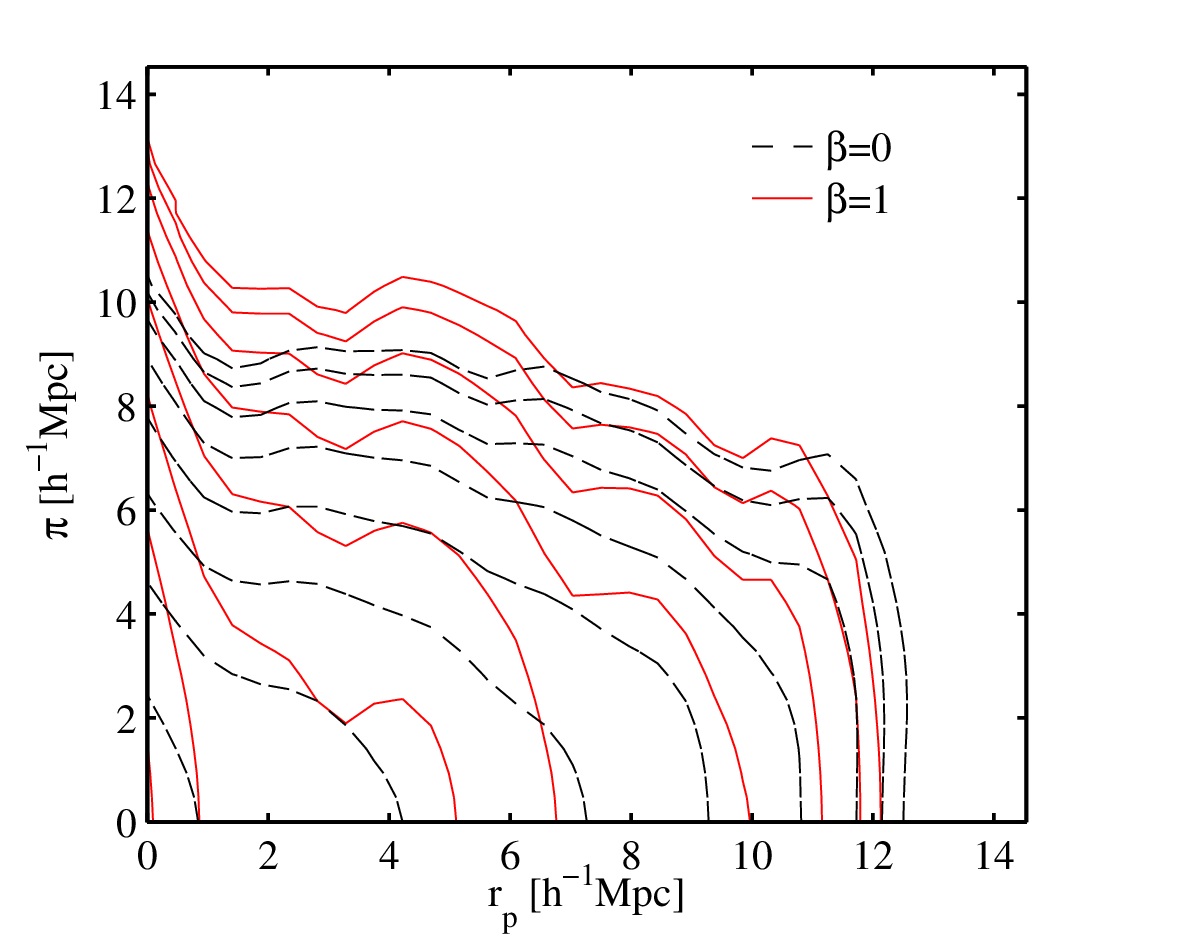, width=8cm,
    height=6.3cm}}
    \caption{Contours of the DM redshift space correlation function $\xi^s(r_{p},\pi)$ as
    function of perpendicular ($ r_p
    $) and parallel ($ \pi $) separations.  The contours are spaced by
    intervals of 0.2 in log($ \xi^s)$, starting with log($ \xi^s)=-1.9$ at the outermost contour.}%
    \label{fig:redshift_dist}
\end{figure}

Galaxy redshift surveys provide redshift distances that differ from real
distances by the radial velocity term, breaking the isotropy of the position
correlation function, thus offering a probe of the peculiar velocity field and
through that the cosmological model. Fig.~\ref{fig:redshift_dist} shows the
redshift space correlation function $\xi^s(r_{p},\pi)$ as a function of the
redshift space separation $r_p$ perpendicular to the line of sight and $\pi$ parallel to the line of sight  \cite{davis83, peebles93},
but here computed for the DM in the standard and \ace cases. 
Random motions on small scales cause the 
spike at $r_p\approx 0$. It is more pronounced for $\beta=1$,
reflecting the stronger small-scale clustering. On scales larger
than $\rs=1\hmpc$, the contours have a similar pattern. We cannot explore the  anisotropy of the redshift space correlation functions of halos and mock HOD galaxies because the anisotropy is sensitive to peculiar
motions and  our simulations are of insufficient dynamical range
to model these motions properly.

\subsubsection{The $\lya$ forest}\label{sec:lyaforest}

The $ \lya $ forest in QSO spectra is caused by resonant scattering of QSO
light by diffuse atomic hydrogen along the line of sight.
This is an important tracer of the distribution of intergalactic baryons at
high redshift  \cite{rauch98}.  The standard $\lcdm$ scenario matches well the
observed one-dimensional (1D) PS derived from the forest
\cite{vince05}.  We present here a preliminary assessment of whether or not
the \ace scenario can also match the 1D PS.  This is an interesting
test because it probes the baryon distribution at epochs where the \ace scenario
significantly differs from the standard model
(Fig. \ref{fig:power}).

The normalized transmitted flux (hereafter flux) in the $\lya$ absorption spectrum is
\begin{equation}
{F}=\exp(-\tau), \qquad  \tau ={\cal A} \rho_{\rm gas}^\alpha.
\end{equation}
The optical depth $\tau$ depends on the mass density $\rho_{\rm gas}$ in baryons, plasma plus neutral, where  $\alpha $ is a parameter in the range $ 1.5 $ to $ 2 $, depending on the photoionization history, and the proportionality
factor ${\cal A}$, which depends on the intensity of the ionizing background
\cite{theuns98}, is given by \cite{nusser00}
\begin{eqnarray}
&&{\cal A} (z)\approx \\
&& 0.61 \;
\left (\frac{300 \kms \mpc^{-1}}{H(z)}\right )
\left (\frac{\Omega_{\rm bar}h^2}{0.02}\right )^{2}\times \ldots \nonumber \\
&&\left ( \frac{\Gamma_{\rm phot}}{10^{-12}\s^{-1}} \right)^{-1}
\left ( \frac{\hat T}{1.5\times 10^{4}\K} \right )^{-0.7}
\left ( \frac{1+z}{4} \right )^{6} . \nonumber
\label{eq:tauf}
\end{eqnarray}
Here $\Omega_{\rm bar}$ is  the baryon density parameter and $\Gamma_{\rm phot}$ is the photoionization rate per hydrogen atom.

We consider the $\lya$ 1D PS of $ F $ for wavenumbers
in the range
\begin{equation}
 0.3 \rhmpc \la  k \la 4 \rhmpc.
 \label{eq:wavenumberrange}
 \end{equation}
 The lower bound is set by fluctuations in the continuum
\cite{hui01} and the upper bound by contamination by metal lines \cite{mcdonald00, kim04}.
We compare the simulations to the observed $\lya$ 1D PS at redshift $z\approx 3$.

To estimate $\rho_{\rm gas}$ as a function of position in the simulations
we interpolate the baryon particle positions by
a cell-in-cloud (CIC) procedure to derive the gas density field on
a $ 256^3 $ uniform cubic grid in the simulation box. This grid
has a Nyquist frequency close to the upper limit beyond which the
flux PS is contaminated by metal lines. The density  field is then
deconvolved by the CIC kernel $ {\rm sinc}^2(k_x L/N_g)\,{\rm
sinc}^2(k_y L/N_g)\,{\rm sinc}^2(k_z L/N_g) $ where $ L $ is the
box length, and $ N_g $ is the number of cells on the side.  The
density field along $ 256^2 $ sight lines is then smoothed with a
1D Gaussian window of width $ \sigma_f $.  This smoothing is
temperature-dependent and is introduced in order to mimic
temperature broadening \cite{vince05}.  It will be treated here
as a free parameter. The resolution of the simulations is
insufficient to model the PS properly at wavenumbers close to
the upper limit in Eq~\ref{eq:wavenumberrange}, so
we degrade the observations to conform with the resolution of our
simulations. We convolve the observed PS with a Gaussian window
\cite{hui01},
\begin{eqnarray}
    P_F(k) \Rightarrow P_F(k) \, \exp{\left(k^2/k_\sigma^2\right)},%
    \label{eq:fwhm}
\end{eqnarray}
where $ k_\sigma=\sqrt{8 \, {\rm ln} \, 2}/\rm{FWHM} $.  We use a FWHM
equivalent to the mean separation between gas particles.  In addition, to
avoid dealing with continuum fitting effects, we truncate the
transmitted flux
\cite{vince05} at $F= 0.8 $.

\begin{figure}[htpb]
    \centerline{\epsfig{figure=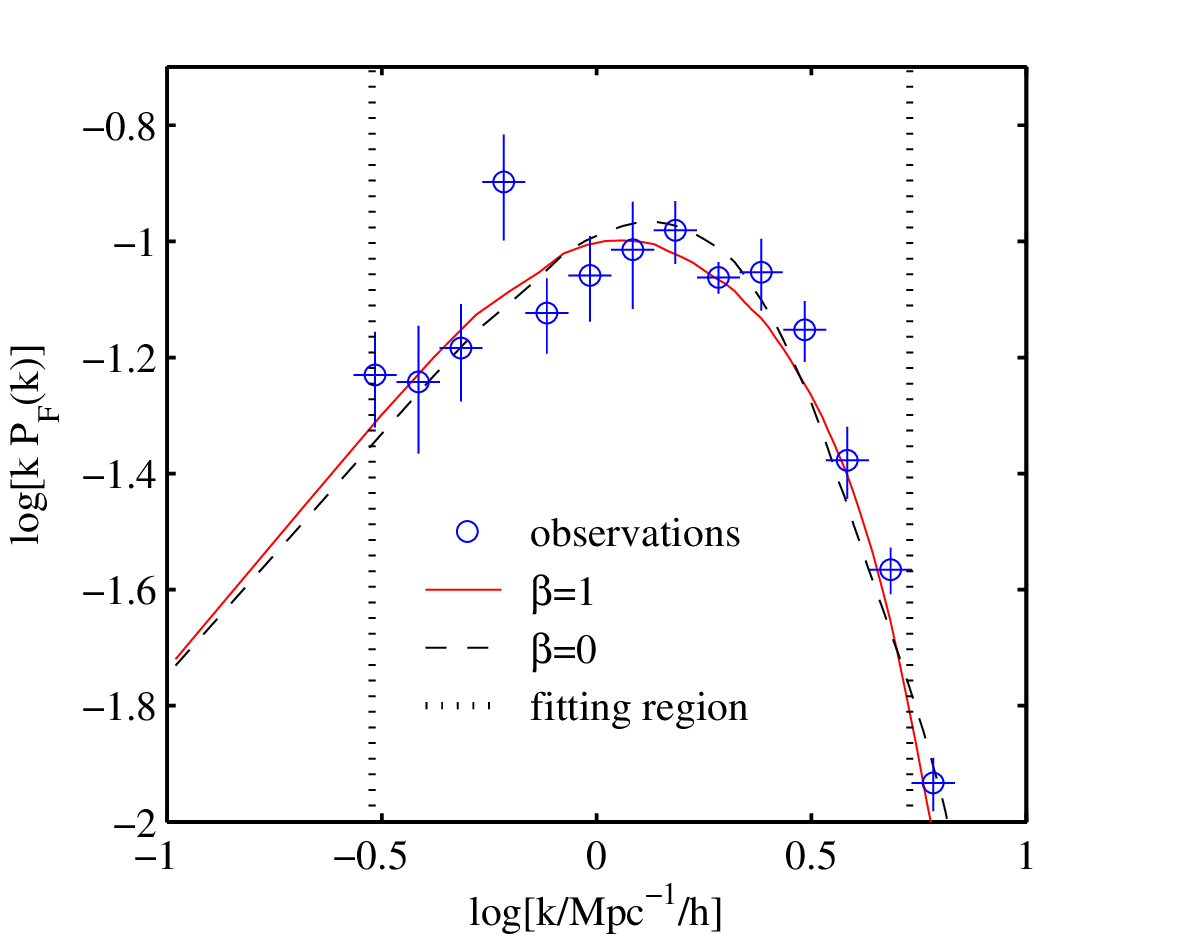, width=8cm,
    height=6.3cm}}
    \caption{The 1D Ly$\alpha$ flux power spectra computed from the $ \beta=0 $ and
    $ \beta=1 $ simulations are shown as the dashed and solid lines
    respectively.  The observed PS is shown as circles
    overlaid with crosses representing the error bars and bin sizes.  The
    vertical dotted lines correspond to the bounds in
    Eq.~\ref{eq:wavenumberrange}.}
    \label{fig:lyman}
\end{figure}

The dimensionless measure $ k\,P_F(k) $ of the observed 1D PS of the $\lya$ transmitted flux compiled in \cite{mcdonald00} is plotted as the open circles in Fig. \ref{fig:lyman}. The curves show the 1D PS computed from the two
simulations. The parameters in these simulated PS
are, for $\beta=0$, $ \alpha=1.99 $, ${\cal A}=0.48$, and $\sigma_f=0.32\hmpc$, and, for $ \beta=1 $,  $ \alpha=1.66$, ${\cal A}=0.71$, and $\sigma_f= 0.35\hmpc $. For mean ICM temperature $\hat T=1.5\times 10^{4}\K$ these parameters give photoionization rates $\Gamma_{\rm phot}=6.9\times 10^{-12} s^{-1}$ for $\beta=0$ and $4.6\times 10^{-12} s^{-1}$ for $\beta=1$.
The values of $ \sigma_f $ correspond to $  k_f\approx 18 \rhmpc $ for the
broadening parameter used in \cite{vince05}, slightly larger than the maximum value adopted in that work.

Fig. \ref{fig:lyman} shows that both $\lcdm$ and \ace can agree with the
observations within the limits of the present analysis. We
caution, however, that this is a preliminary test because our
simulations neither include a direct treatment of photoheating and
photoionization nor they are of sufficient resolution to resolve
the Jeans mass.

\subsubsection{Clustering of galaxies}
\label{sec:clusteringofgalaxies}

Galaxies form in DM halos where the gas density can reach high
enough values to cool, contract, and form stars.
Therefore, in contrast to the $ \lya $ forest, which traces the baryons,
the distribution of galaxies is dictated by the DM density field.

\begin{figure}[htpb]
    \centerline{\epsfig{figure=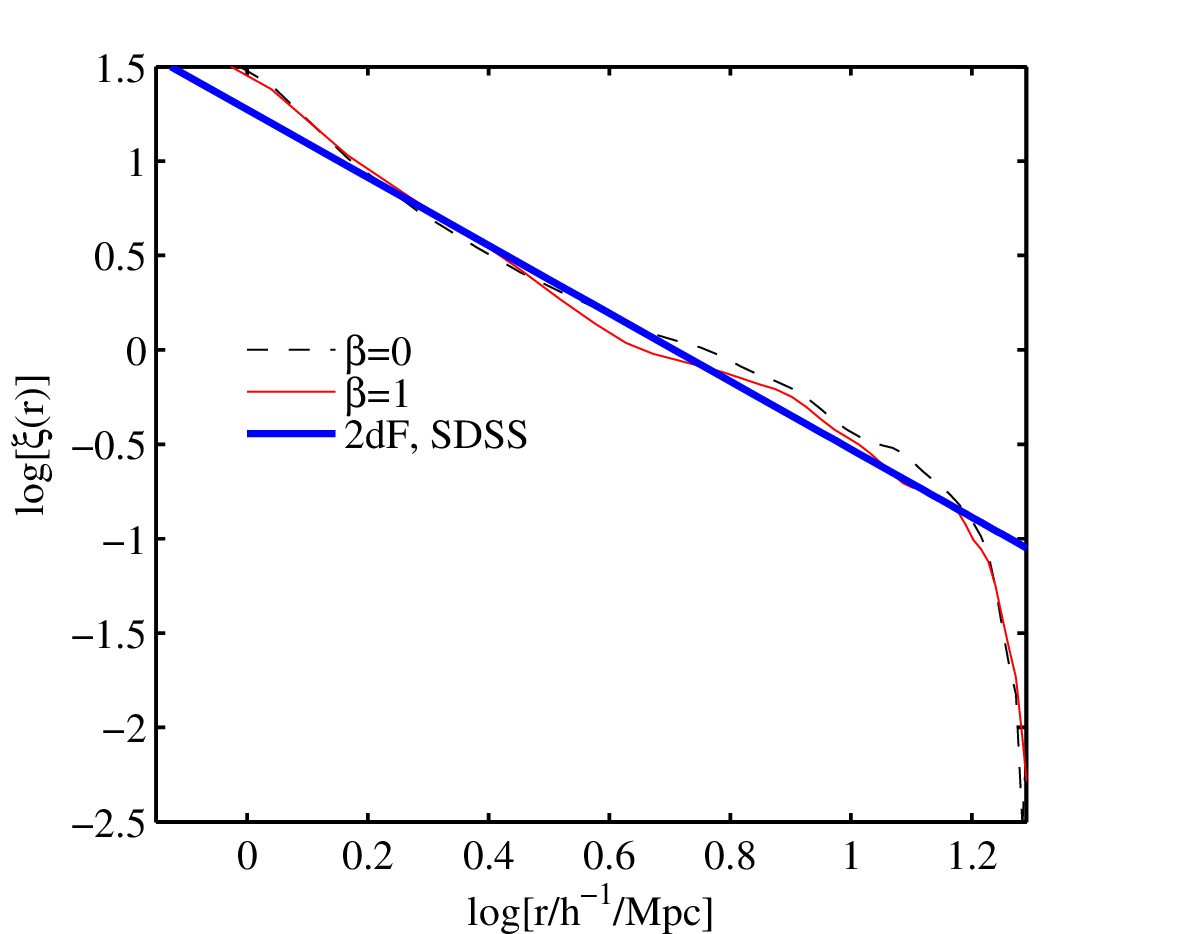, width=8cm,
    height=6.3cm}}
    \caption{Correlation functions for the positions of the HOD-placed galaxies in
    the $ \beta=0 $ and $ \beta=1 $ simulations are shown as the thin
    dashed and thin solid lines respectively.  The
    observed correlation function of galaxies is derived from the 2dF and
    SDSS surveys.}%
    \label{fig:gal_corr}
\end{figure}

In the simulations we populate halos with ``galaxies" according to the HOD framework described in \S~\ref{sub:g}. The HOD parameters for the $ \beta=0 $ simulation are $\alpha=1.13 $, $ M_{_{\rm min}}=3 \times 10^{11} $, and $ \msat=8\times 10^{12} \hms $.  They are similar to the ones
derived in \cite{zehavi05}.  The parameters for the $ \beta=1 $
simulation are $ M_ {_{\rm s}} \simeq 1.3 \times 10^{13}\hms $, $
\alpha=1.13 $ (as for $ \beta=0 $), $ \gamma=1.8 $, $ M_{_{\rm
min}}=2.2 \times 10^{12} $, $ \msat=6.3 \times 10^{12} \hms $.
In the $\beta=1$ simulation the multiplicative factor, $\zeta$,
by which satellite distances from the halo center are increased
relative to the DM distribution in the host halo is $\zeta=4$.

\begin{figure}[htpb]
    \centerline{\epsfig{figure=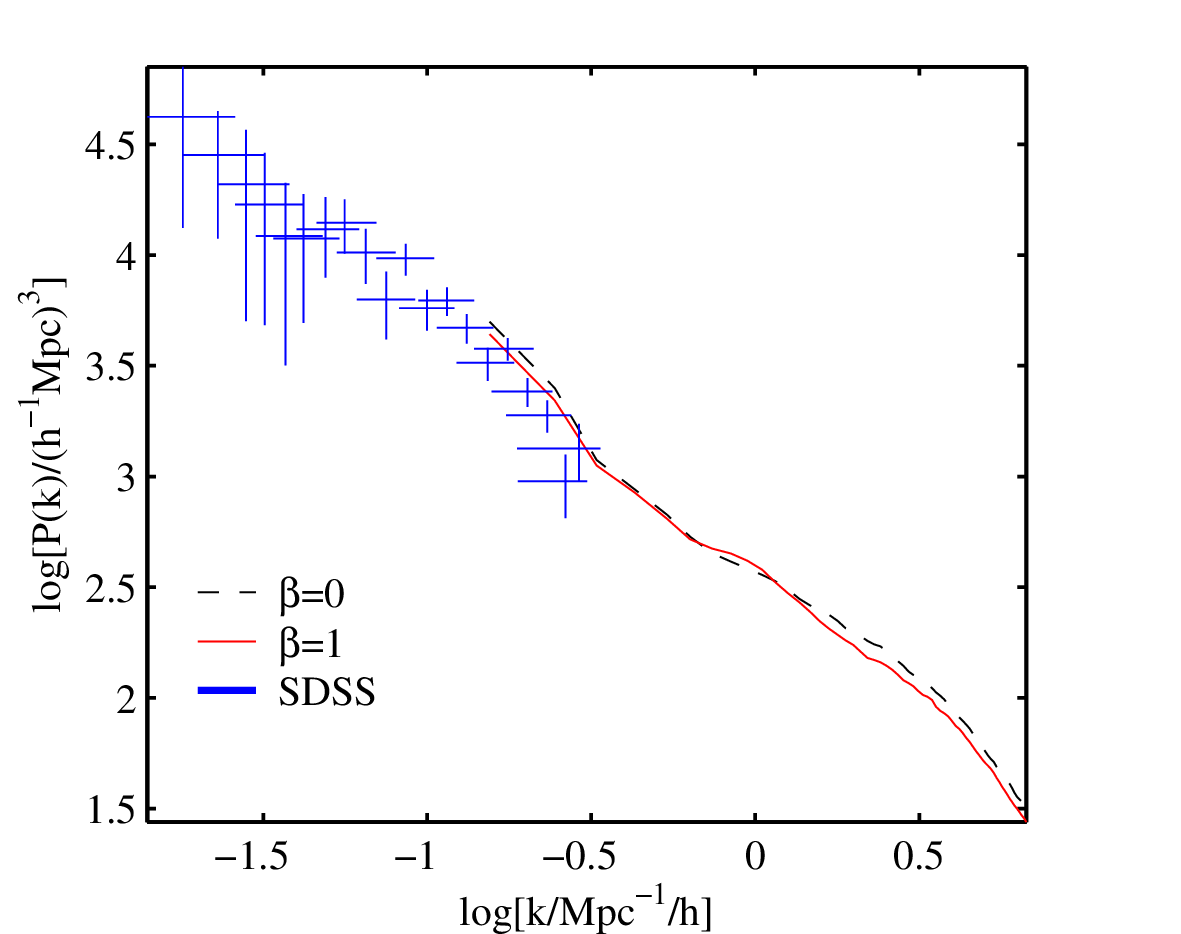, width=8cm,
    height=6.3cm}}
    \caption{Galaxy PS from HOD galaxies in the simulations and in the SDSS
    galaxy survey.}%
    \label{fig:gal_ps}
\end{figure}

The  solid and dashed curves in Fig. \ref{fig:gal_corr}
show the correlation functions of these galaxies
in the two simulations.  The thick solid line
is the fit $ \xi=(r/5.1 \hmpc)^{-1.8} $ to the observed
correlation function extracted from Fig. 11 in
\cite{krav04}. In the range of separations shown here the correlation
functions in the two simulations, with the adopted HOD parameters,
agree reasonably well with each other with what is observed. The
alternative view from the galaxy PS is shown in Fig.~\ref{fig:gal_ps}.
The observed galaxy PS is taken from the SDSS galaxy redshift survey
(table 2 of \cite{tegmark04}).

An important test not examined here is the shape of the correlation
function of the simulated galaxies on smaller scales, where \ace may
be expected to have a larger effect. That will require simulations
with better resolution. 

In the range of galaxy separations we can explore the fit
of the position correlation function of the \ace simulation
HOD galaxies to the observed galaxy correlation requires HOD
parameter $\zeta=4$. That is, \ace requires that the
distribution of satellite galaxies is considerably more
extended than the DM in the host halo.
This is in contrast to the standard cosmology in which
only a mild bias, or none at all (as in this work), is
needed to match the galaxy correlation function. In
high resolution simulations of  standard $\lcdm$ the
distribution of the DM sub-halo population is more extended than
the distribution of diffuse DM in the host halo
\cite{ghigna98, colin99, ghigna00, springel01, delucia04, gao04}.
That is, HOD satellite galaxies
which trace the general DM distribution cannot be associated
directly with the sub-halo population. We may expect that sub-halos in
the \ace scenario are also more extended than the DM
distribution in the host halo, but that remains to be checked, along with
the possibility that satellites galaxies could be more directly
related to sub-halos in the \ace scenario.

\begin{figure}[htpb]
    \centerline{\epsfig{figure=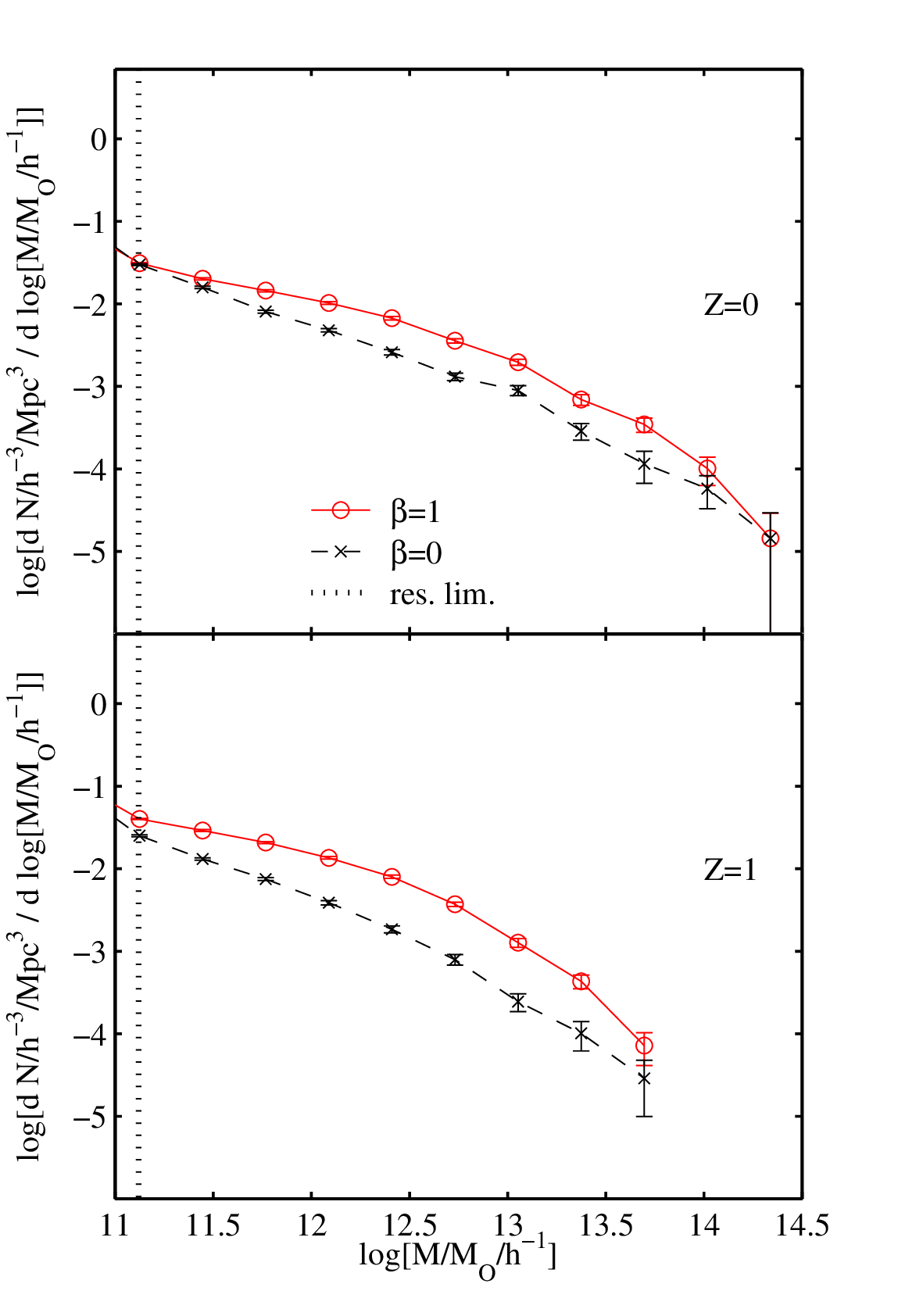, width=8cm,
    height=11.41cm}}
    \caption{Mass function of halos in the $ \beta=0 $ and $ \beta=1 $
    simulations at redshifts $ z=0 $ and $ z=1 $, for DM particles only.
    The vertical dashed line marks a mass of a halo containing 20 DM
    particles as identified by the FOF algorithm.}%
    \label{fig:halohist}
\end{figure}

\subsection{Properties of halos}

\subsubsection{The halo mass function}\label{sec:halomassfunction}

To explore the abundance of DM halos as a function of their
mass we select FOF groups containing more than $ 20 $ DM
particles each, giving a minimum halo mass of $ 1.3\times 10^{11}
\hms $.  The mass functions expressed as the number density of
halos per logarithmic mass bin are plotted in Fig.
\ref{fig:halohist}.  The $ 1\sigma $
error bars are estimated by the Bootstrap resampling procedure described in \cite{keselman07}.

The mass functions in the two simulations are nearly the same at
$ M\approx 10^{11}\hms $, which is close to the minimum halo mass
we can identify. The difference between mass functions also is small
at $M\approx 10^{14}\hms$, about the mass at $z=0$ at which the virial
radius is equal to the screening length $\rs=1\hmpc $. At the still larger
masses characteristic of clusters of galaxies the \ace force is less
effective and we may expect the mass function to be less sensitive
to $\beta$. Our computation box size is too small to explore this.

The mass functions in the \ace and standard $\lcdm$ simulations at
the present epoch differ by a factor of about two at
$M\sim 10^{12}\hms$, roughly the DM mass of the Milky Way \cite{dehnen06}.
Translating this to the predicted effect on the galaxy luminosity function does not seem likely to be reliable at the present state of the art.

The difference between the mass functions is greater at redshift $z=1$,
and the mass function expressed in comoving
coordinates evolves more slowly at $z<1$ in the \ace scenario.
As we discuss next, this agrees with the earlier completion of merging in \acec.

\begin{figure}[htpb]
    \centerline{\epsfig{figure=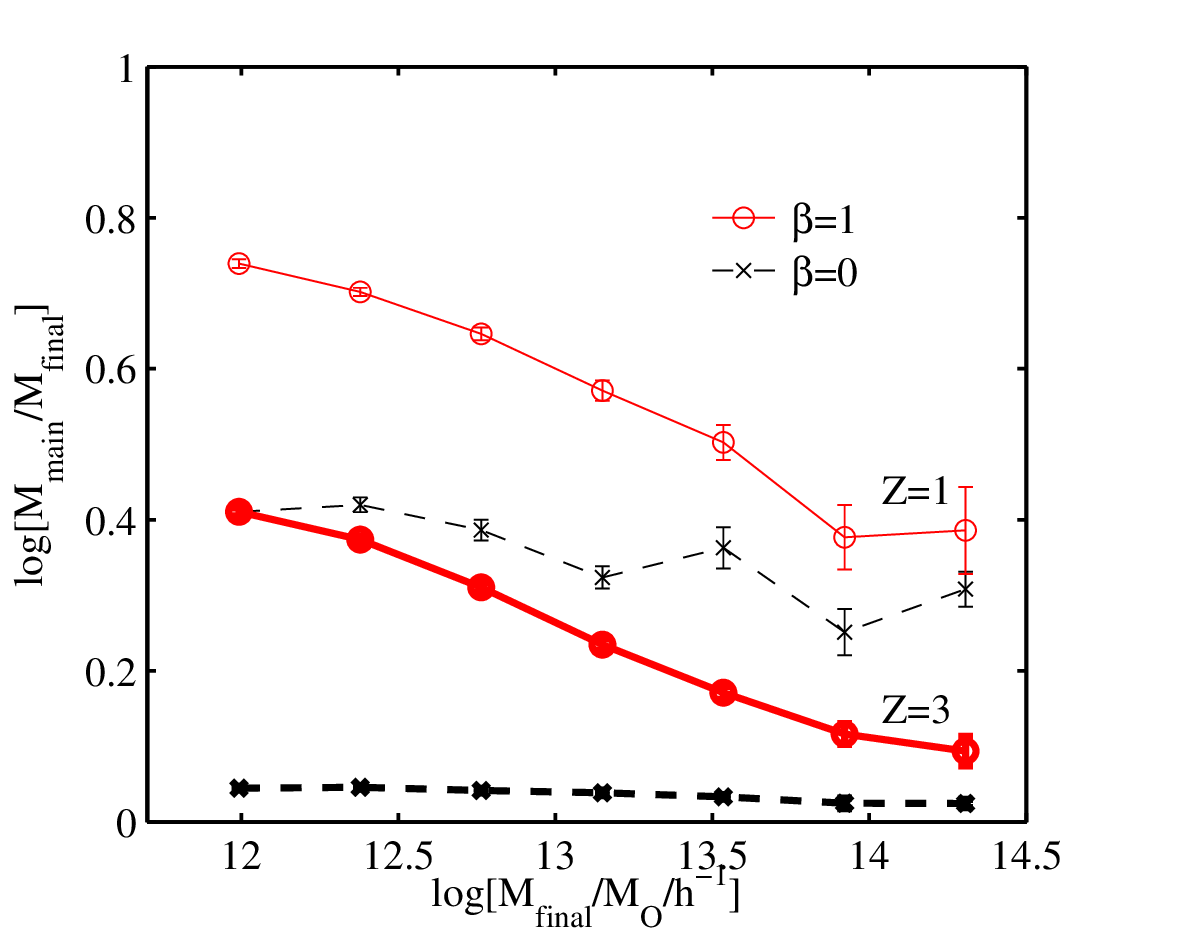, width=8cm,
    height=6.3cm}}
    \caption{Mean fractions of mass assembled by redshifts $ Z=1 $
    and $ Z=3 $ as functions of the present halo mass in the simulations
    with and without \acec.}%
    \label{fig:mergers}
\end{figure}

\subsubsection{Halo Merging History}\label{sec:halomerginghistory}

A present-day halo went through a time of intense merging activity
when it acquired a significant fraction of its present mass. For a
measure of when this  happened we identify in each DM halo at $z=0$
the particles belonging to the most massive progenitor (MMP) at
redshifts $ z=1 $ and $ z=3$. Fig. \ref{fig:mergers} shows the average
ratios of the mass of the MMP to the present halo mass, $\mfin$, as a
function of $ \mfin $. The $ 1\sigma $ error-bars are estimated by
Bootstrap re-sampling.

The figure indicates that  $z=1$ a halo with the present mass
$10^{12} \hms $ typical of the Milky Way has assembled on average 75\%
of its mass in the \ace model with $ \beta=1$, and about half that in
the standard model with  $\beta=0$. At $z=3$ a halo now typical of
the Milky Way has assembled 40\% of its mass in the \ace model, while
at $\beta=0$ the MMP typically contains fewer than the limiting 20
particles needed for the FOF halo identifications. Similar results
are reported by \cite{hellwing09}.
This effect is strongest for halos with masses typical of galaxies
and it nearly disappears at masses typical of rich clusters, where
it will be recalled that the virial radius exceeds the screening
length $\rs$. As discussed in \S~\ref{sec:sec_issues}, the earlier
assembly of DM halos in \ace with $\beta=1$ may play an important
role in resolving issues of galaxy formation.

\subsubsection{Halo Mass Profiles}\label{sec:halomassprofile}

Here we consider the DM mass density
as a function of radius in halos at $z=0$ that satisfy the convergence
criteria \cite{power03} that the two-body relaxation time within
the studied radius is much longer
than the Hubble time, the smoothing length of the particles does
not allow two-body accelerations to be larger than the mean
field acceleration, and there are at least 50 particles within the innermost radius studied. That leads us to examine halos containing
each at least 2500 particles identified in the FOF algorithm.
For each halo we calculate the spherically averaged mass profile
centered on the most bound particle in the halo. Here and in 
what follows we present the mean density $\rho(<r)$ of mass  
within radius $r$, rather than the density at radius $r$, to suppress 
the noise from the limited numbers of particles. 

\begin{figure}[htpb]
    \centerline{\epsfig{figure=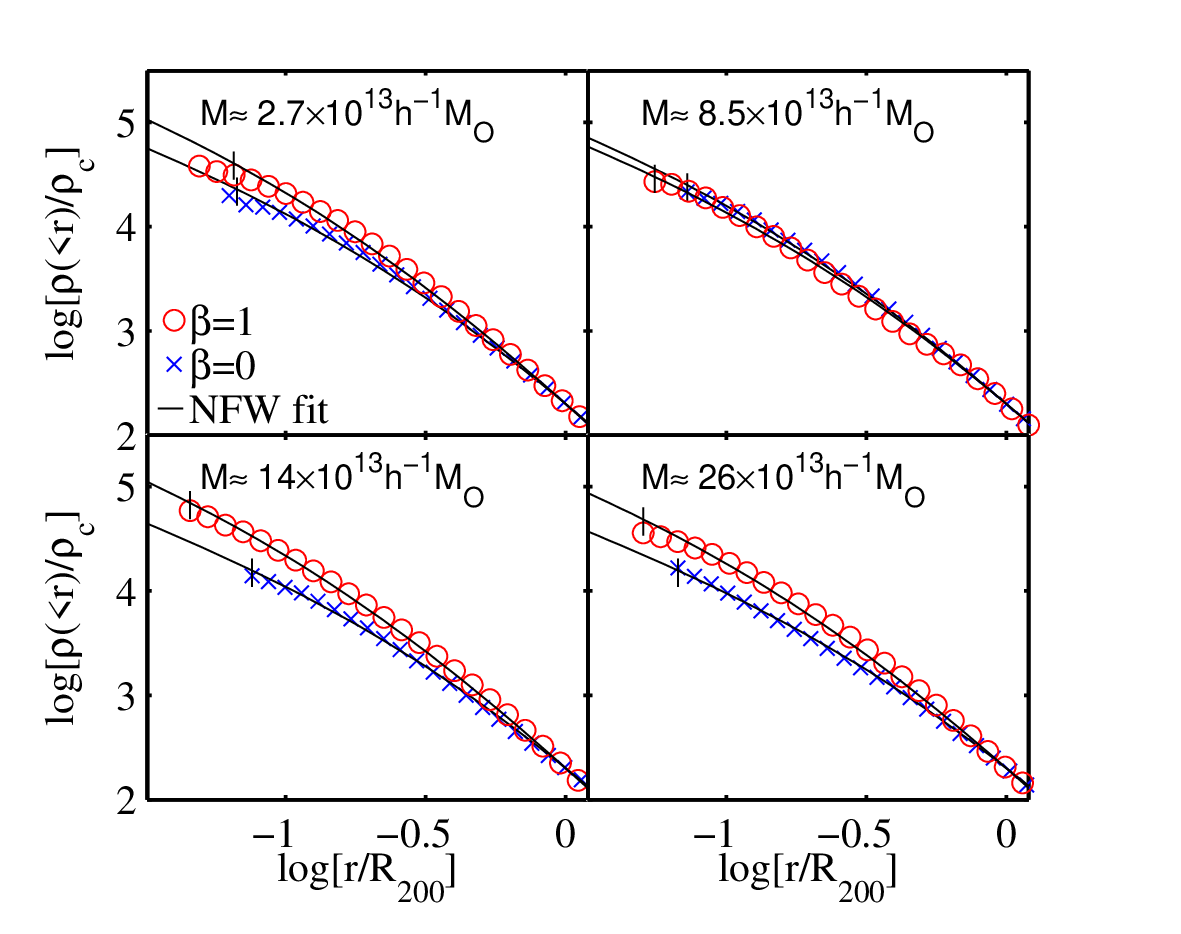, width=8cm,
    height=6.3cm}}
    \caption{Mass density runs at $z=0$ in both
    simulations for four randomly chosen halos at the
    indicated masses.  The short vertical lines mark the innermost radii that
    could be studied within the convergence requirements. The curves are
    fits to the NFW form.}%
    \label{fig:nfw_fit}
\end{figure}

Mean density profiles measured for four  halos in each simulation are shown in Fig. \ref{fig:nfw_fit}. The inner vertical lines attached to each profile indicate the minimum radius, $r_{\rm{min}}$, above which the convergence criteria are satisfied.

The curves in Fig. \ref{fig:nfw_fit} are integrals of the NFW form \cite{nfw97}
\begin{equation}
\rho(r) = \frac{\rho_{_{0}} \rtwo^3}{\cnfw r (\rtwo+\cnfw r)^2}.
\label{eq:NFW}
\end{equation}
For each halo in the figure we compute the radius $ \rtwo $ within which the mean halo density is $ 200 $ times the critical density, $\rhoc$. Then the concentration parameter $ \cnfw $ is adjusted to fit the mass density run (where $ \rho_{_{0}} $ is required to make the  mean density
within $\rtwo$ equal to  $200\rhoc$).

We use two measures of the performance of the NFW profile under a
\ace force. The first is a `Fit Error' parameter
\begin{eqnarray}
   {\rm Fit~Error} \propto  \log \left[ \frac{1}{\triangle R}\int \, \left( \log{\rho / \rho_{_{\rm
    {NFW}}}} \right) ^2 \, \rm{d}r \right].
    \label{eq:quality}
\end{eqnarray}
The integration is from $r_{\rm{min}}$ to $ \rtwo $, $\rho $ is the measured density within $ r $, $ \rho_{_{\rm{NFW}}}
$ is the fitted NFW density, and $ \triangle R \equiv
\rtwo-r_{\rm{min}} $. The second measure is a `Relaxed' parameter.
\begin{equation}
{\rm Relaxed} = \log{\rtwo/r_{_{\rm{cm}}}} \label{relaxed}
\end{equation}
where $ r_{_{\rm{cm}}} $ is the
distance of the center of mass of the FOF-group from its
most bound particle.  The more spherical the halo the
larger its  `Relaxed' parameter.

\begin{figure}[htpb]
    \centerline{\epsfig{figure=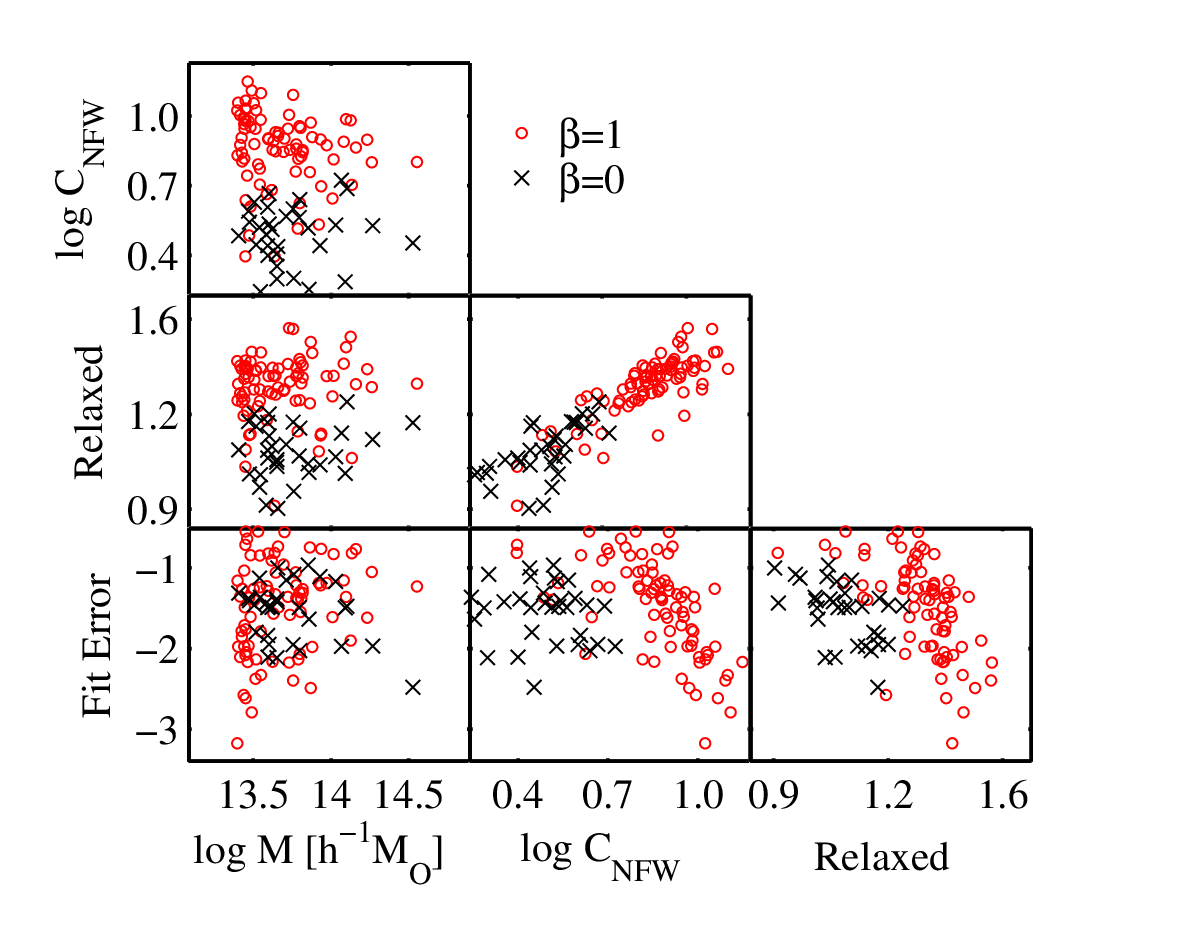, width=8cm,
    height=6.3cm}}
    \caption{Correlations of halo properties for standard and
    \ace model halos.}%
    \label{fig:halo_corr}
\end{figure}

Fig.~\ref{fig:halo_corr} assesses the  correlations between pairs of these
halo structure parameters at $z=0$. Circles and crosses
correspond to $\beta=1$ and $\beta=0$. The three panels in the
bottom row are scatter plots of  the `Fit-Error' parameter versus the
halo mass $M$, the concentration parameter $\cnfw$ in Eq.~\ref{eq:NFW},
and the `Relaxed' parameter in Eq.~\ref{relaxed}. These
panels indicate the NFW profile about equally well matches the halo mass
profile in both simulations. This agrees with the findings of
\cite{hellwing09}, who used a different measure than Eq.~\ref{eq:quality}.
These scatter plots also make the possibly important point that the \ace
scenario produces halos that are more relaxed and more concentrated.

\begin{figure}[t]
    \centerline{\epsfig{figure=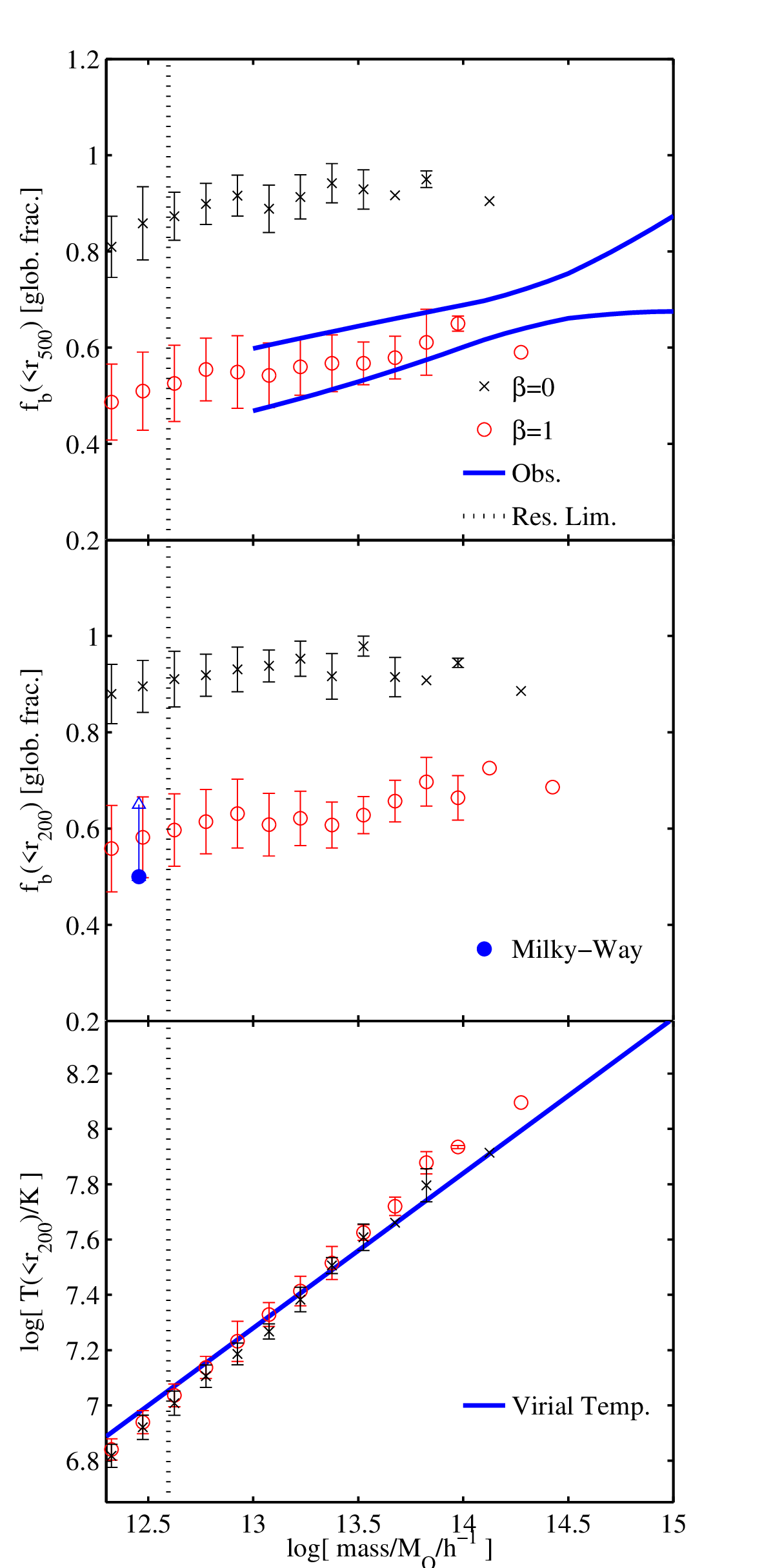, width=8cm,
    height=15.7cm}}
    \caption{Baryonic mass fraction $f_b$ versus DM halo mass.
    Top and  middle  panels, respectively,  correspond to masses
    within $\rfive$ and $\rtwo$. Solid curves in the top panel are
    $\pm1\sigma$ limits on the mass fraction from recent observations.
    The bottom panel shows the mean gas temperature within $\rtwo $ versus the
    DM halo mass within $\rtwo$; the solid line is the virial temperature relation.}%
    \label{fig:ratio}
\end{figure}

\subsubsection{Baryonic mass fraction in halos}
\label{sec:baryonmassfraction}

The baryon mass fraction $f_b$ is the ratio of the
mass density in baryons to the mass density in baryons
plus dark matter, normalized to unity at the cosmic mean
$\Omega_b/\Omega_m$. We consider here the measured \cite{mcg08, mcg09} and computed values of $f_b$ in galaxies and systems of galaxies. 

In the largest clusters of galaxies $f_b$ is close to unity \cite{zhang06, gonz07}; these systems are tightly enough bound that they have resisted large loss 
of baryons. The heavy blue solid curves in the top panel in Fig. \ref{fig:ratio} are the $\pm1\sigma$ limits on observed baryon fractions in lower mass halos.
This includes the mass in X-ray emitting plasma,
and the mass in stars, as  given in parametric form in \cite{giodini09}.
Two adjustments are made to these components.
First, the stellar mass is increased by $11\%$ to take account of the
contribution from
intracluster light \cite{zibetti05, krick07,murante07}.
Second, $f_b$ is reduced by
$10\%$, as a method to overcome the systematic errors arising in our simulations,
which do not model radiative losses and star formation processes,
which could increase the mass fraction in the simulated haloes by $10\%$
\cite{kravtsov05, ettori06}. Baryonic energy dissipation through radiation
increases the total baryonic fraction in haloes because gas flows
deeper into haloes to replace cooled gas.

The top and middle panels in Fig. \ref{fig:ratio} show the
computed means of the baryon mass fractions $f_b$ within distances 
$\rfive$ and $\rtwo$ from the most bound particle in halos at $z=0$ in our
simulations with $\beta=0$ (black crosses) and $\beta=1$ (red circles).
The horizontal axis is the DM halo
mass within $\rfive$ for the top panel and $\rtwo$ for the center panel.
The vertical dotted lines mark the mass in a halo containing a minimum mass
equal to the sum of $ 500 $ DM and $ 500 $ baryonic particles,
the threshold for reliable estimation of mass fractions in simulations
suggested by 
the convergence studies of \cite{crain07,naoz09}. The computed $f_b$ in our
standard $\lcdm$  simulation agrees with previous studies
\cite{crain07, gottl07}. We find that the variation of $f_b$
with redshift is weak in both the $\beta=1$ and $\beta=0$ simulations
(c.f. \cite{crain07, navarro95, eke98, ettori06, kravtsov05}). 

The bottom
panel of the figure shows the  mean gas temperature within $\rtwo$ as a
function of the DM mass within $\rtwo$. The solid line is computed from
the virial relation ($T\sim G M/\rtwo$). Halos in the \ace (circles) and
standard (crosses) simulations both closely follow the virial relation.

The top panel of Fig. \ref{fig:ratio} illustrates the well-established
evidence that  in galaxies and groups of galaxies roughly half the
baryons are missing. At halo mass $\sim 10^{12} M_\odot$ the plasma
temperature is $\sim 10^6$\,K, cool enough to escape detection as an X-ray
source. In galaxies at this mass and lower the missing baryons could be in
the DM halo as a galactic corona  \cite{FukugitaPeebles06}. At larger mass
the missing baryons have to have been removed  from their DM halos, perhaps
by galactic superwinds driven by supernovae and active galactic nuclei
\cite{efs00, puch08, bower08, CenOstriker06}. The \ace scenario adds
another possibility, that differential acceleration of baryons and DM has
separated some baryons from their halos. 

The weight of evidence for or against \ace from the baryon mass fraction $f_b$
in galaxies and systems of galaxies depends on the quite uncertain astrophysics
of baryon winds. If these processes were effective in suppressing $f_b$ then
\ace could make $f_b$ unacceptably small. If winds proved to be ineffective it
would argue for \ace at about the parameter values for the DM force we have
adopted, for the observed and computed baryon fractions agree in our \ace
simulation. The \ace prediction for clusters, which have close to the cosmic
baryon fraction, requires larger simulations. The blue filled circle in the
middle panel in Fig. \ref{fig:ratio} shows that the observed baryon fraction
in the Milky Way is well below the $\lcdm$ case and agrees with \acec. This
is particularly interesting because the relatively low mass of the central
black hole in our galaxy might argue against removal of the baryons by a
superwind. But the Milky Way is at about the mass threshold where the missing
baryons may be in a corona cool enough to be detected as an X-ray source.

\begin{figure}[htpb]
    \centerline{\epsfig{figure=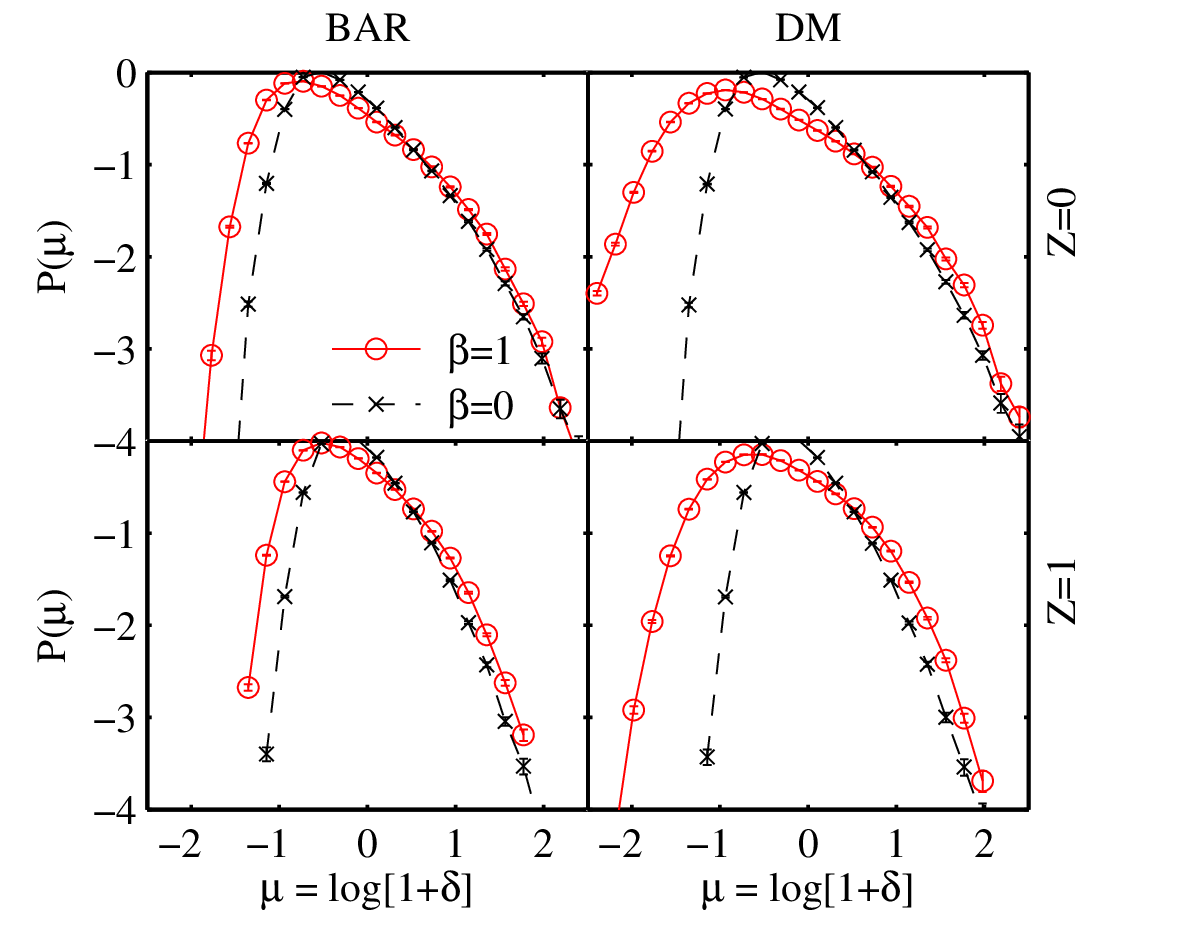, width=8cm,
    height=6.3cm}}
    \caption{Probability distributions of the densities in baryons and DM 
    smoothed by a top-hat window of radius $ 1.5\hmpc $}
    \label{fig:pdf}
\end{figure}

\subsection{Void properties}
\label{sub:sub_voids} 

 Fig. \ref{fig:pdf} shows the effect of \ace on the probability
distribution functions of the mass densities in baryons and DM
averaged within randomly placed top-hat windows of radius $1.5\hmpc$.
The densities are normalized to the cosmic means at the computed
redshifts, $z=1$ and $z=0$, as the density contrast $\delta=\rho/{\bar \rho}-1$. \ace more efficiently evacuates low density regions, as previously demonstrated \cite{nusser05}. Here
we explore in more detail the effects of \ace on voids.

We define a void in the simulation as a spherical
region of radius $R$ with average density contrast 
$\delta_{\rm v}$. We identify voids in the distribution of
particles in a simulation as follows.
We derive a density field by CIC interpolation of the
points to a cubic grid with 128 cells on the side.
The density field is then smoothed with a 
Top-Hat window with radius large enough that there are no grid points 
with density contrast less than $\delta_{\rm v}$. The width
of the Top-Hat window is then gradually decreased until a single
grid point with $\delta =\delta_{\rm v}$ is encountered. This grid
point is taken to be the center of the largest void and its size is the
window radius $R$. The filtering radius is further
decreased to locate successively smaller voids, excluding those overlapping
with the larger, already identified, voids \cite{cec06}.

\begin{table}[htpb]
\caption{Counts of Voids}
\begin{tabular} {c c c c }
\hline\hline
$\beta$\quad &\qquad $\delta_v$\qquad &$4<R< 6$\footnotemark[1] \  &\ $6<R<9$\footnotemark[1] \\
\hline
0 &\quad $-0.8$ & 36 & 5\\
1 &\quad $-0.8$ & 54 & 16\\
0 &\quad$-.09$ &   4 & 0 \\
1 &\quad$-.09$ &  69 & 6 \\
\hline
\end{tabular}
\footnotetext[1]{unit $= \hmpc$}\label{table:voids}
\end{table}

Table \ref{table:voids} lists counts of voids in the $\lcdm$ and \ace simulations
for two ranges of values of the void radii and the threshold density contrast
$\delta_v$. \ace considerably increases the numbers of voids, as also illustrated
in Fig.~\ref{fig:pdf}, and the effect on the number of larger voids is even more
pronounced at the lower threshold $\delta_{\rm v}= -0.9$. The number of smaller
voids is smaller at $\delta_{\rm v}= -0.8$ because many of the small voids at
$\delta_{\rm v}= -0.9$ become parts of larger voids at $\delta_{\rm v}= -0.8$.

\begin{figure}
    \centerline{\epsfig{figure=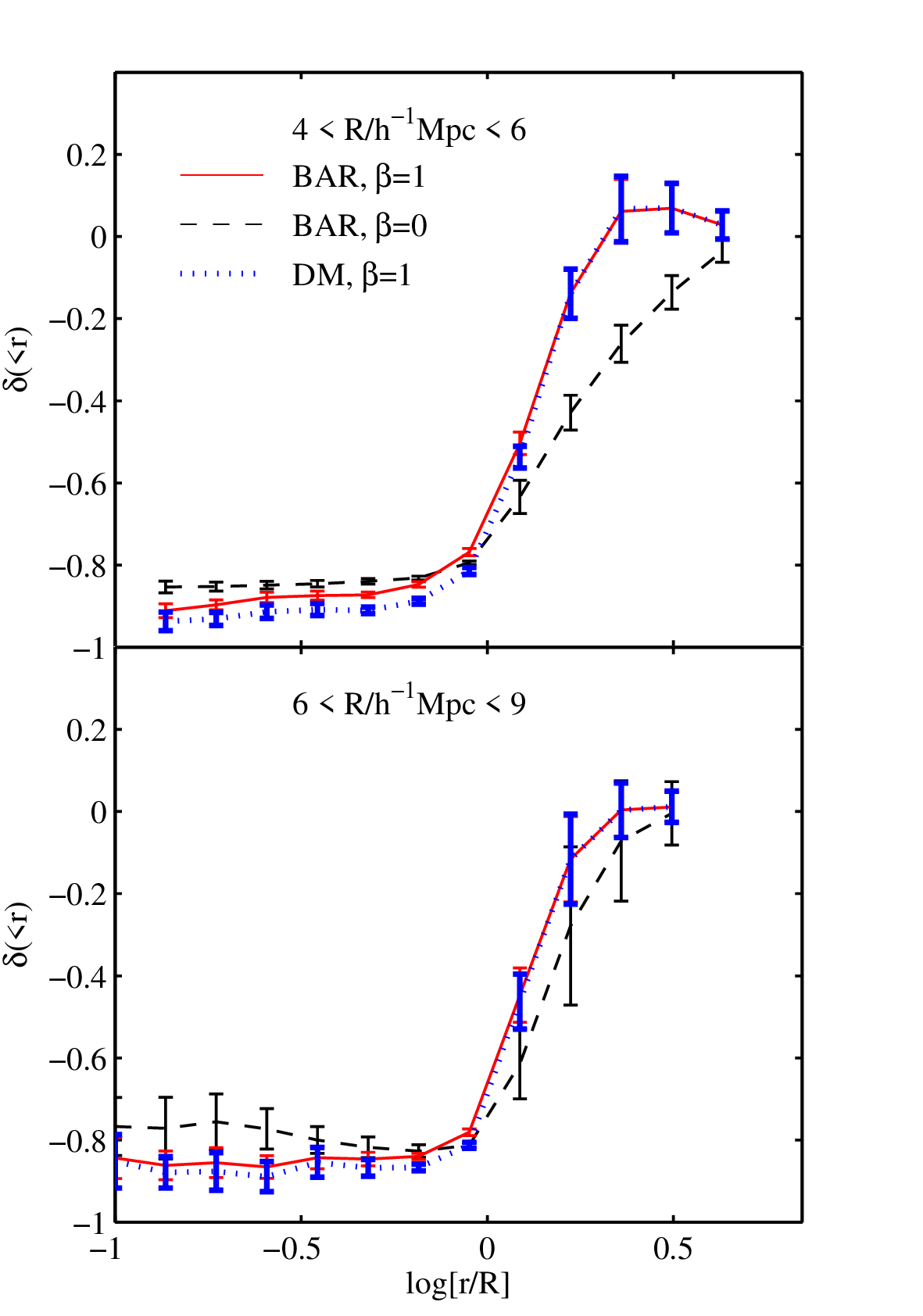, width=8cm,
    height=11.41cm}}
    \caption{Void mean density within radius $r$ at redshift $z=0$ for large (bottom) and small voids
    (top). The void radius is $R$.}%
    \label{fig:voids}
\end{figure}

We represent the mass distribution within a void as the density
contrast computed from the mass contained within distance $r$ from its center.
This measure is less noisy than the local mass density profile.
Fig.~\ref{fig:voids} shows means of the void profiles for small (top panel) and
large voids (bottom). \ace reduces the DM density in the central
regions and produces a steeper density gradient at the void edge.
Both effects are more pronounced in smaller voids, consistent with
\cite{martino09}. This could ease the tension between the
observations that indicate that small void profiles have a sharper
transition into the outer regions than what appears in $\lcdm$
simulations \cite{cec06}.

The DM particle mass in our simulations is too large to allow an
analysis of the distribution of low mass halos in voids. We can
note, however, that the distinct suppression of the mean DM density
in voids is in the wanted direction to suppress the numbers of void
galaxies, while the lesser suppression of baryons may be more relevant
for the density of diffuse plasma in voids. 

\begin{figure}[htpb]
    \centerline{\epsfig{figure=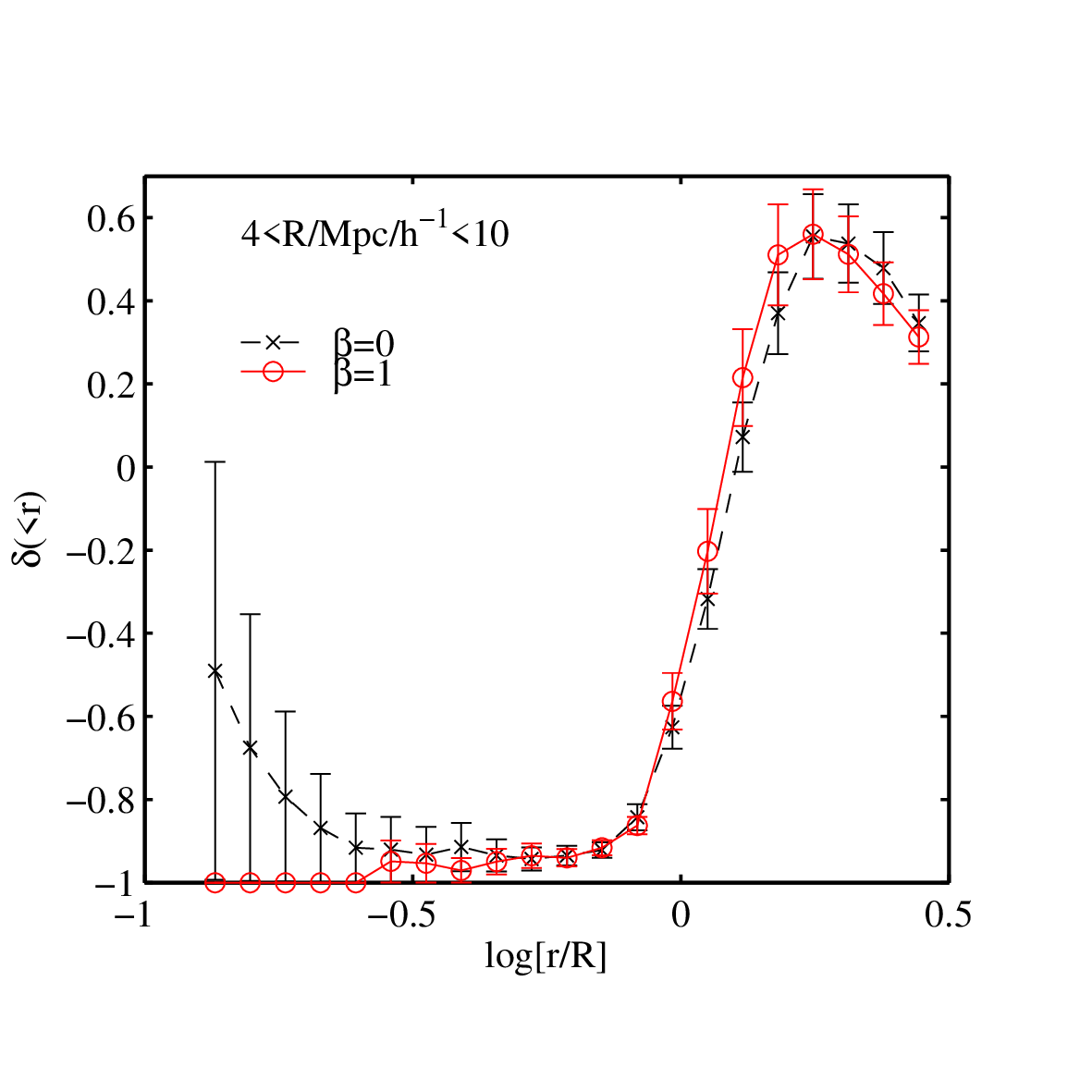, width=8cm,
    height=6.3cm}}
    \caption{Mean density contrast of HOD galaxies in voids.}%
    \label{fig:void_gal_prof}
\end{figure}

\begin{figure}[htpb]
    \centerline{\epsfig{figure=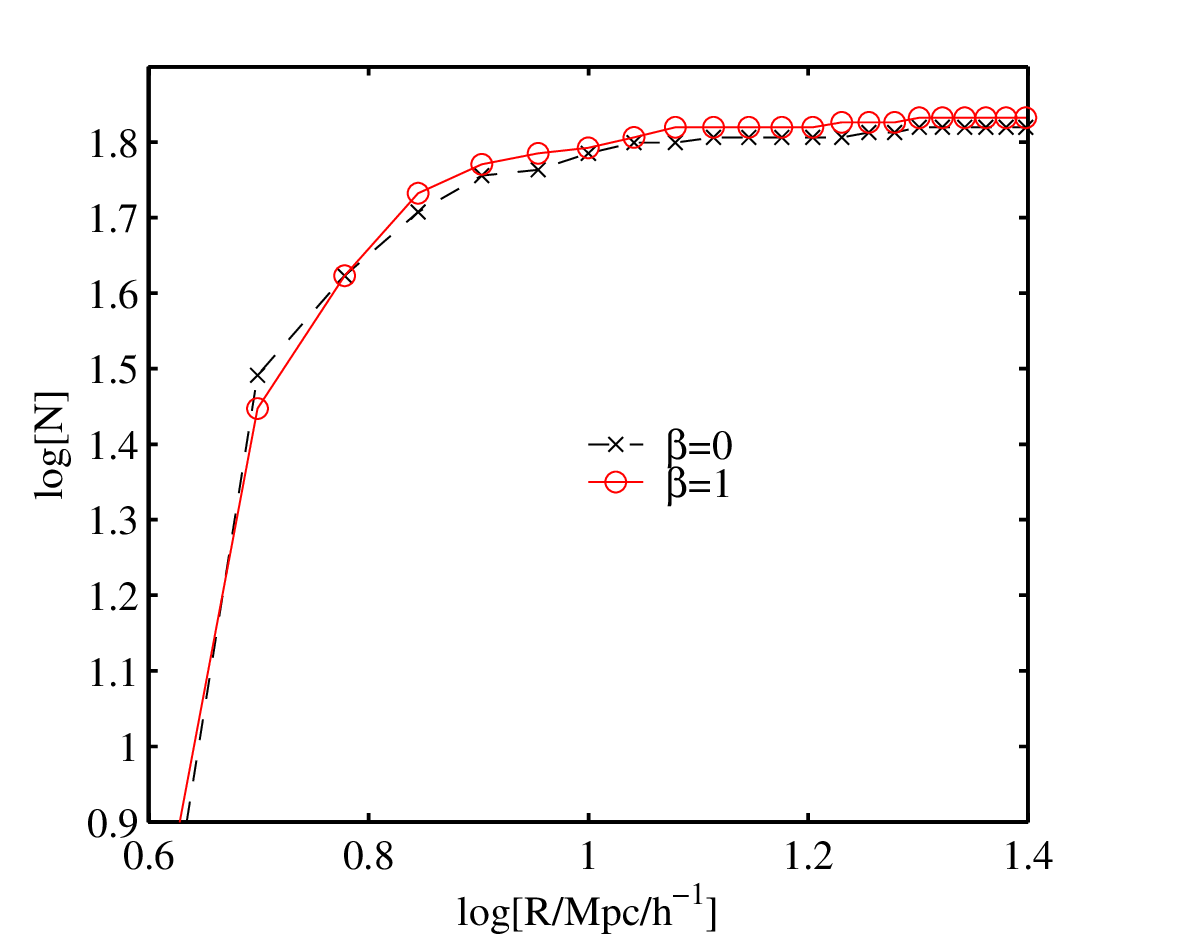, width=8cm,
    height=6.3cm}}
    \caption{Number of voids smaller than $ R $ and larger than $ 4\hmpc
    $.}%
    \label{fig:void_gal_hist}
\end{figure}

Figs. \ref{fig:void_gal_prof} and \ref{fig:void_gal_hist} show
checks that ReBEL can match the void size distribution and the
distributions of bright galaxies in voids as well as the standard
model. The galaxies are modeled for absolute magnitude limit $-19.5$
in the $r$ band. The HOD parameters were
tuned so that similar void properties for these galaxies are
obtained in \ace and the standard cosmology. The vertical axis in
Figs.~\ref{fig:void_gal_prof} is the mean HOD galaxy number averaged
within radius $r$ and scaled to the void radius $R$.
Fig.\ref{fig:void_gal_hist} shows the number of voids in our
simulations as a function of the void radius, using $\delta_v=-0.8$.
We conclude from the consistency of results for
$\beta =0$ and $\beta=1$ in the statistics in
Figs. \ref{fig:void_gal_prof} and \ref{fig:void_gal_hist} that at
the level of our simulations the bright galaxy distribution does
not a usefully discriminate between $\lcdm$ and \acec.

\section{DISCUSSION}%
\label{sec:sec_discussion} 

We have considered the effect of \acec, a long-range force of attraction
acting on the dark matter alone. The force is modeled by Eq.~\ref{eq:eq1}
with $\beta=1$, meaning the strength is the same as gravity, and screening
length $\rs=1\hmpc$, meaning it  has no influence on the cosmological tests
but can have interesting effects on galaxies. 

\ace with the parameters adopted here satisfies the known constraints
from properties of satellites of the Milky Way, as discussed elsewhere
\cite{frieman93, kk06, keselman09}, but this certainly requires closer
analysis. Within the framework of the Halo Occupation Distribution
(HOD) model for galaxies we have shown that \ace fits the measured
galaxy correlation function at separations greater than about one
megaparsec as well as does the standard $\lcdm$ cosmology. Checking
galaxy clustering on smaller scales will require larger simulations. 

We have also concluded that \ace can account for the properties of the
Lyman-Alpha forest about as well as $\lcdm$, at the level of semi-analytic
methods \cite{vince05, bi92}. Again, this requires deeper analysis, including
resolution of the Jeans mass of the intergalactic medium and an explicit treatment of 
photoionization and photoheating by the ionizing radiation background. 

Fig. \ref{fig:ratio} shows that \ace offers a natural explanation of the
missing baryons in galaxies and groups of galaxies. This may be important
because the hypothesis that superwinds removed the baryons from isolated
galaxies and loose groups does not have a secure theoretical or empirical basis. 

Fig.~\ref{fig:mergers} shows that \ace substantially advances the redshift
of assembly of the bulk of the mass of a galaxy. This may be important in
resolving the puzzle of disk-dominated galaxies \cite{Croft09}. In $\lcdm$
inefficient star formation allows development of dwarf pure disk galaxies
\cite{Governato09}, but that approach would make the disk of the Milky Way
unacceptably young \cite{Wyse09}. \ace would help by assembling the parts
of a galaxy earlier, and assembling a lower baryon mass fraction. Much
better resolution simulations will be needed to explore  whether \ace can
promote formation of significant numbers of bulge-free spiral galaxies
with acceptable galaxy rotation curves, along with appreciable numbers of
large red galaxies at high redshifts. 

Earlier structure formation may also be relevant to the measurements of
polarization of the cosmic microwave background radiation that 
indicate a significant reionization by redshift
$z\sim 11$, implying early structure formation. Star
formation in the $\lcdm $ with $\sigma_8\approx 0.9$ could be
early enough to account for such  early reionization, but 
extreme efficiency of UV photon production by early structures is
needed for $\sigma_8\approx 0.8 $, the value favored by  WMAP data
\cite{benson06}. \ace may relieve this condition.

Earlier assembly is accompanied by more massive DM halos at the present
epoch. The effect is not large, however, as illustrated in Fig.~\ref{fig:halohist},
and it would seem to be a difficult challenge to translate this into a test of
\ace from the galaxy luminosity function. 

Figure~\ref{fig:voids} shows the depression of the DM mass density within voids,
and the lesser suppression of the baryon mass density. This is in the wanted
direction of resolving the void over-population problem, but larger simulations
will be needed to estimate the effect of the lower DM density on the formation
of low mass DM halos. Finally, we illustrate in Fig.~\ref{fig:halo_corr} the
formation of larger halo NFW concentrations in \acec, which is in the direction
suggested by recent  measurements \cite{broadhurst05, umetsu08, broadhurst08, broadhurst05b, lemze08}.

\section*{ Acknowledgement} This work was supported by THE ISRAEL SCIENCE
FOUNDATION (grant No.203/09), the German-Israeli Foundation for Research
and Development, the Asher Space Research Institute and by the WINNIPEG
RESEARCH FUND. AN is grateful for the Princeton Institute for Advanced
Study for the hospitality.

%\bibliography{refs}
\bibliography{KNP09B.bbl}

\appendix

\section{Modification of Gadget2}%
\label{app:app_modification} For the simulation with $ \beta=0 $ we used
the original Gadget2 public version. For the simulations with $ \beta=1 $,
we modified the original code as follows: in the PM part of
the code we added a second loop on top of the original.  This loop
evaluates the density of the DM component and transforms it into the
scalar potential by convolution with the proper green
function, given in $ k $ space by
\begin{equation}
    G(k)=-\frac{4\pi G \beta}{k^2+r_{s}^{-2}}.%
    \label{eq:eq20}
\end{equation}
The tree part of the code was modified so that every tree node would
include a center of charge (scalar charge, equals DM mass times $ \beta $)
in addition to the center of mass. A cell-opening criterion was developed,
such that the relative error in the sum of the \ace and gravitational
forces is consistent with the original relative error of the gravitational
force. The \ace force is added in a similar way to the gravitational one,
that is, by considering the first moment.

To constrain the scalar
force relative error we consider the worst-case scenario, this 
is when we calculate the force on a DM test particle, located co-linearly 
with two DM particles at the opposite corners of a
cubical cell of width $L$. The relative error is of course independent of
the global interaction strength $ \beta $ but not on the scale
length $ \rs $. To first non-vanishing order in $ L/r $, where $r$ is the distance of
the test particle from the center of mass/charge of the cubic cell, and up to a
factor of $ 4/3 $ the relative error is given by
\begin{equation}
    e_s=\frac{\left(r^3+3 r^2 \rs+6 r \rs^2+6 \rs^3\right) L^2}{6 r^2
    \rs^2 (r+\rs)}%
    \label{eq:eq21}
\end{equation}
which goes back to the usual gravitational relative error $ e_g=L^2/r^2 $
for $ \rs \gg r $. Since the particle is given a scalar force and a
gravitational one, the total relative error is constrained by
\begin{equation}
    e_t=\frac{e_g F_g+e_s F_s}{F_g+F_s}%
    \label{eq:eq22}
\end{equation}
which then gives the new opening criterion as
\begin{equation}
    e_t^n \left( F_g^n+F_s^n \right) < \alpha \left( F_g^{n-1}+F_s^{n-1}
    \right)%
    \label{eq:eq23}
\end{equation}
where $ e_t^n $, $ F_g^n $, and $ F_s^n $ are evaluated at the $ n $th
time-step and $ \alpha \ll 1 $ is the bounding relative error per time
step.

\section{Simulation tests}%
\begin{figure}[htpb]
    \centerline{\epsfig{figure=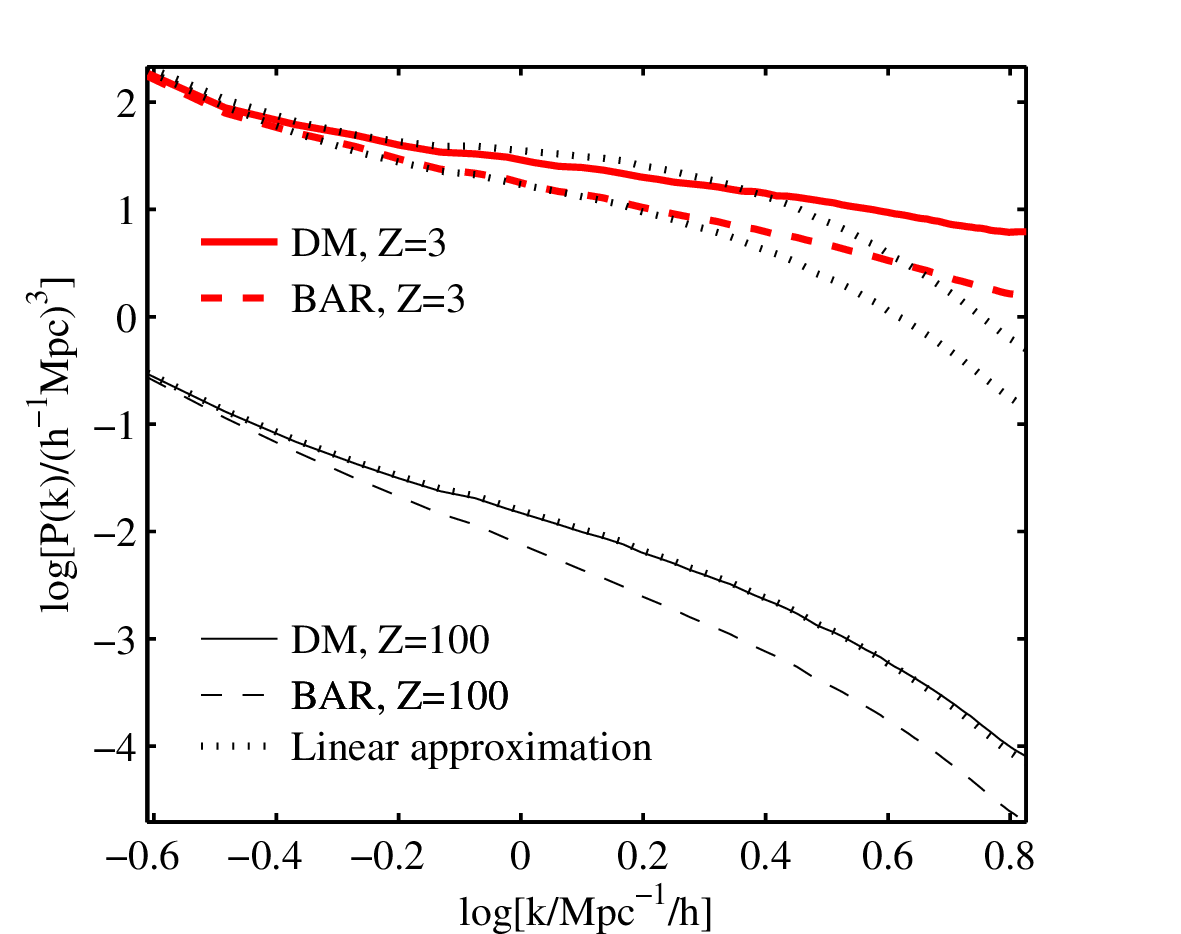, width=8cm,
    height=6.3cm}}
    \caption{Simulation PS compared with linear theory, as described in
    appendix \ref{app:app_linear}, for $\beta=1$.}
    \label{fig:linear}
\end{figure}

\label{app:app_tests} The modified code was checked at the large,
cosmological scales, by comparing simulations to the \ace linear theory
with a baryonic component included (as described in App. \ref{app:app_linear}).
This is illustrated in Fig. \ref{fig:linear},
where the linear approximation is calculated for the
baryonic component PS at redshift $ Z=100 $ and $ Z=3 $, and for the DM
component only at the latter.  These calculations use no more than the
measured PS of the DM at $ Z=100 $, a time when the decreasing solutions
have almost vanished.  The linear approximation seems to fit very well
the simulations at the linear scales.  At smaller scales the modified
code was checked by two more tests, as described in
\cite{keselman09}. The $ \beta=0 $ simulation was checked at all scales
by comparing the PS shown in Fig. \ref{fig:power}
to the PS of the corresponding GIF simulation in
\cite{smith03}.

\section{Linear Theory}%
\label{app:app_linear}
First we write the fundamental equations of fluid motion in physical
coordinates.  These consist of the Euler equations,
\begin{eqnarray}
    \frac{{\rm d} \ub}{{\rm d} t} + \frac{\nab {\rm P}}{\rbar} + \nab
    \phibar + \nab \phidm &=& 0%
    \label{eq:euler_full_bar}\\
    \frac{{\rm d} \ud}{{\rm d} t} + \nab \phibar + \nab \phidm + \nab
    \phis &=& 0,
\end{eqnarray}
where $ \ub $ and $ \ud $ are the baryonic and DM velocity fields, and $
{\rm d}/{{\rm d} t} $ is the usual convective derivative, and is equal
to $ \partial /{ \partial t}+{\bf u} \cdot \nab $. The symbols $ \phibar
$, $ \phidm $, and $ \phis $, represent the baryonic, DM, and scalar
potential.  These are given by the poison equations
\begin{eqnarray}
    \nab^2 \phibar - 4 \pi {\rm G} \rbar &=& 0\\
    \nab^2 \phidm - 4 \pi {\rm G} \rdm &=& 0\\
    \nab^2 \left( \phis - \rs^{-1} \right)- 4 \pi {\rm G} \beta \rdm &=&
    0
\end{eqnarray}
where the last equation is a screened poison equation describing the
scalar potential, and $ \rs $ is a function of time, and is proportional
to the scale factor $ a(t) $. Since this study deals with scales much
larger than the Jeans scales, and the simulations are dissipation-less,
we neglect the baryonic pressure and radiative cooling terms.  Thus, we
skip the baryonic equations of energy and state, which play no role in
the structure formation dynamics, and conclude this presentation with
the continuity equations:
\begin{eqnarray}
    \frac{{\rm d}\rbar}{{\rm d}t}+\rbar \nab \ub &=& 0\\
    \frac{{\rm d}\rdm}{{\rm d}t}+\rdm \nab \ud &=& 0.
\end{eqnarray}

We define $ \delta \ub $ as the baryonic peculiar velocity and $ \delta
\pot $ as the dark matter peculiar gravitational potential. In addition,
it is assumed that the baryonic mass fracture is much smaller than that
of DM (hence the gravitational potential is given only by the DM). Under
these assumptions, the linearised Euler and Continuity equations, for
the baryons, in physical (Eulerian) coordinates may be written as
\cite{peacock_book}
\begin{eqnarray}
    &\delta \dub = - \nab \delta \pot - H \delta \ub \\
    &\ddg = -\nab \cdot \delta \ub.
\end{eqnarray}
where $ \delta \equiv \delta \rbar / {\bar \rho}_{bar} $, and $ \delta
\rbar $ is the peculiar baryonic density. From now on these equations
will be developed working in comoving coordinates.  To achieve this, we
transform $ \nab \rightarrow \nab / a $ and $ \delta \ub \rightarrow a
\vg $, where $ a $ is the scale factor, to get
\begin{eqnarray}
    &\dvg = - \nab \pot - 2H \vg\\
    &\ddg = -\nab \cdot \vg
\end{eqnarray}
where $ \nab \pot $ is given by the correspondent poison equation. In $
K $-space $ \nab \rightarrow i \bk $ and $ \nab^2 \rightarrow -k^2 $ so
in transforming these equations to $ K $-space we have
\begin{eqnarray}
    &\dvg = - i\bk \pot - 2H \vg%
    \label{eq:euk1} \\
    &\ddg = -i\bk \cdot \vg%
    \label{eq:contk1}
\end{eqnarray}
Eq. \ref{eq:euk1} is then multiplied by $ i\bk $ and
Eq. \ref{eq:contk1} is differentiated relative to $ t $ to get
\begin{eqnarray}
    &i\bk \cdot \dvg = k^2 \pot - i\bk 2H \vg%
    \label{eq:euk2} \\
    &\dddg = -i\bk \cdot \dvg%
    \label{eq:contk2}
\end{eqnarray}
taking into account Eqs. \ref{eq:contk1} and \ref{eq:contk2} we have
\begin{eqnarray}
    -\dddg = k^2 \pot + 2H \ddg%
    \label{eq:euk3}.
\end{eqnarray}
The poison equation of gravity in comoving coordinates is
\begin{eqnarray}
    \pot = -\frac{4 \pi G \mrdm \dm}{a^3 k^2}= -\frac{3H^2\om\dm}{2 k^2}
    \label{eq:pois3}
\end{eqnarray}
with $ \om \equiv \mrdm \rho_{crit}^{-1} a^{-3} $ and $ \rho_{crit}
\equiv 3 H^2 / 8\pi G $. Hence, Eq. \ref{eq:euk3} can be written as
\begin{eqnarray}
    \dddg + 2H \ddg = \frac{3}{2}H^2\om\dm%
    \label{eq:euk4}
\end{eqnarray}
and it is useful to define $ \om $ in terms of it's current value $ \omz
$ as $ \om = \omz H_0^2 H^{-2} a^{-3} $. in a similar way we can derive
the linear equation for perturbations in the dark-matter density, except
that this time the total potential that a DM particle feels is the
gravitational potential plus the scalar potential, giving rise to a
modified poison equation
\cite{nusser05}
\begin{eqnarray}
    \phi_k = \pot \left(1+\frac{\beta}{1+(\bk r_s)^{-2}}\right)
\end{eqnarray}
and so all in all, the perturbation equation is
\begin{eqnarray}
    \dddm + 2H \ddm = \frac{3}{2}H^2\om C \dm%
    \label{eq:euk5}
\end{eqnarray}
where
\begin{eqnarray}
    C \equiv 1+\frac{\beta}{1+(\bk r_s)^{-2}}
\end{eqnarray}
and
\begin{eqnarray}
    H^2&=&(\frac{\dot a}{a})^2 \\
    &=&H_0^2[\omz(1+z)^3+\olz+\Omega_0(1+z)^2] \nonumber
\end{eqnarray}
where the curvature parameter is $ \Omega_0 \equiv \omz+\olz $. Taking $
z \equiv 1/a-1 $ to high numbers, one can see that $ \om \rightarrow 1 $
and the density fluctuations behave as in an Einstein-DeSitter universe.

for a constant $ r_s $ in an Einstein De-Sitter universe ($ \Omega=1 $
and $ \Omega_{\Lambda}=0 $) $ a \propto t^{2/3} $ and $ H \propto t^{-1}
$ it can be checked by inspection
\cite{nusser05} that $ \dm \propto t^p $ where
\begin{eqnarray}
    p=\frac{1}{6} \left( 25+\frac{24\beta}{1+(\bk r_s)^{-2}} \right)^{\frac
    {1}{2}}-\frac{1}{6}
\end{eqnarray}
in contrast to the $ p=2/3 $ factor when no scalar interactions are
present. As for the baryonic fluctuations, $ \dg = \dm / C $.

In this study we generalize the results of
\cite{nusser05} for arbitrary $ \omz $ and $ \olz $, while using
regular scaling relations.  For commodity reasons, from now on wherever we
refer to $ \delta(\om,C) $ if either $ \om $ or $ C $ equal 1 they will
be omitted.  So for example, the notation of $ \delta(\om=1,C) $ becomes
$ \delta(C) $ and $ \delta(\om=1,C=1) $ becomes just $ \delta $.  In
addition, when no species is defined for $ \delta $, it may be either
one of them (either $ \delta_{dm} $ or $ \delta_{bar} $).

\begin{figure}[htpb]
    \centerline{\epsfig{figure=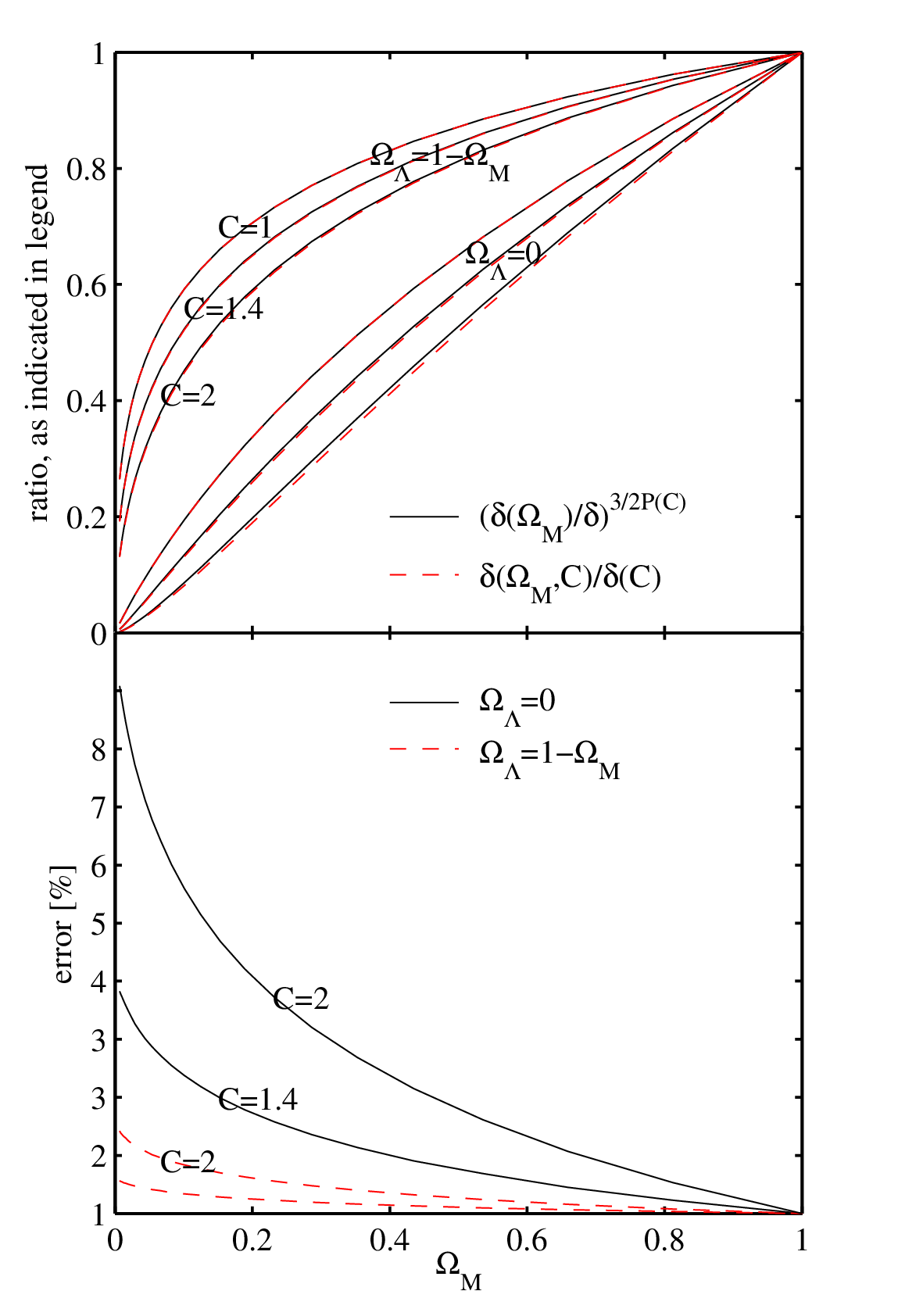, width=8cm,
    height=11.41cm}}
    \caption{Approximate analytic solution to linear theory.  Upper
    panel compares the numerical calculation in dashed red lines, to the
    approximation, which corresponds to the solid blue lines.  The
    comparison is made both for a flat universe and for an universe
    without dark energy, for different values of $ C $, as indicated in
    the figure.  The bottom panel shows the relative error in the
    approximation.}%
    \label{fig:approx}
\end{figure}

We claim by numerical integration, that a good approximation for the
solution of Eqs. \ref{eq:euk4} and \ref{eq:euk5} is
\begin{eqnarray}
    \frac{\delta(\om,C)}{\delta(C)} \simeq \left(\frac{\delta(\om)}{\delta}\right)^
    {\frac{3}{2}P}%
    \label{eq:res1}
\end{eqnarray}
This can be seen in Fig. \ref{fig:approx}.
The relative error of this approximation is quite
small, as can be seen in the lower panel of the same figure.

for the case of $ C=1 $ we have $ \beta=0 $ and there are no scalar
interactions - thus we may use standard approximations for the base of
the RHS. for example
\cite{carroll92}
\begin{eqnarray}
    \frac{\delta(\om)}{\delta} \simeq \frac{\frac{5}{2}\om}{\om^{4/7}-\olm+
    (1+\frac{1}{2}\om)(1+\frac{1}{70}\olm)}.
\end{eqnarray}
Since $ \delta(C) \propto a^{\frac{3}{2}P} $ and $ \delta \propto a $ we
can see from%
\ref{eq:res1} that $ \delta(\om, C) \propto \delta(\om)^{\frac{3}{2}P} $.
And so we get an important result for velocity fields, namely
\begin{eqnarray}
    f \equiv \frac{{\rm d} \log(\delta(\om,c))}{{\rm d} \log(a)} \simeq
    \frac{3}{2}P \frac{{\rm d} \log(\delta(\om))}{{\rm d} \log(a)}
\end{eqnarray}
and the derivative term of the RHS can be approximated in the usual
form, for example
\cite{peebles93}
\begin{eqnarray}
    \frac{{\rm d} \log(\delta(\om))}{{\rm d} \log(a)} \simeq \om^{0.6}
\end{eqnarray}
this is almost independent of $ \Lambda $, as shown in
\cite{lahav91}, which also gives a better approximation for a flat
universe:
\begin{eqnarray}
    \frac{{\rm d} \log(\delta(\om))}{{\rm d} \log(a)} \simeq \om^{0.6}+\frac
    {1}{70} \left( 1-\frac{1}{2}\om(1+\om) \right)
\end{eqnarray}

Next we define the baryonic velocity potential in $ K $-space as $ i\bk\phiv
\equiv \vg $ so Eq. \ref{eq:contk1} is
$ \ddg = k^2 \phiv $.  Noting that $ \ddg=f\dg H $ we
then have
\begin{eqnarray}
    \frac{3}{2}Pf_1\dg H = k^2 \phiv%
    \label{eq:pot2}
\end{eqnarray}
where $ f_1 $ is the normal factor $ f $, with $ C=1 $.  This result is
of course good also for the DM fluctuations.  The velocity-density
relation is thus
\begin{eqnarray}
    \frac{3}{2}Pf_1\dg H = -i \bk \cdot \vg.
\end{eqnarray}
Linear theory is further extended by giving the baryons mass.  In this
case, Eq. \ref{eq:euk5} becomes
\begin{eqnarray}
    \dddm + 2H \ddm = \frac{3}{2}H^2(\omdm C \dm+\omg \dg).
    \label{eq:barmass1}
\end{eqnarray}
Lets assume that $ \dg=\dm/f $ with constant $ f $.  Then the last
equation becomes
\begin{eqnarray}
    \dddm + 2H \ddm = \frac{3}{2}H^2\omdm C^* \dm%
    \label{eq:barmass2}
\end{eqnarray}
where $ C^* \equiv \omg/(f \omdm)+C $.  The solution to this equation
was already given before.  As for the baryons, Eq. \ref{eq:euk4} becomes
\begin{eqnarray}
    \dddg + 2H \ddg & = \frac{3}{2}H^2(\omg \dg+\omdm \dm)%
    \label{eq:barmass5}
\end{eqnarray}
Then, using the definition of $ f $, this becomes
\begin{eqnarray}
    \dddm + 2H \ddm & = \frac{3}{2}fH^2(\omg \dm/f+\omdm \dm) \\
    & = \frac{3}{2}fH^2 \omdm (\omg /\omdm+f)\dm%
    \label{eq:barmass6}
\end{eqnarray}
which is true if and only if it becomes Eq. \ref{eq:barmass2}. That means,
\begin{eqnarray}
    \frac{\omg}{f \omdm}+C=\frac{\omdm}{\omg}+f,
    \label{eq:barmass7}
\end{eqnarray}
which gives
\begin{eqnarray}
    &&f = \\
    &&\frac{C\omdm-\omg+\sqrt{(C\omdm-\omg)^2+4\omdm\omg}}{2\omdm},
    \nonumber%
    \label{eq:barmass8}
\end{eqnarray}
and indeed this factor is constant, since $ \omdm $ and $ \omg $ have
the same dependence on time.

This theory is compared to numerical simulations, as seen in
Fig. \ref{fig:linear}, and as described in App. \ref{app:app_tests}.
This comparison indicates on the correctness of
both the simulations and the linear theory.

\end{document}